\documentstyle[12pt]{article}

\def\hybrid{\topmargin -20pt  \oddsidemargin 0pt
      \headheight 0pt   \headsep 0pt
      \textwidth 6.25in 
      \textheight 9.5in 
      \marginparwidth .875in
      \parskip 5pt plus 1pt   \jot = 1.5ex}

\hybrid

\begin{document}
\def\x{\times}
\def\be{\begin{equation}}
\def\ee{\end{equation}}
\def\bea{\begin{eqnarray}}
\def\eea{\end{eqnarray}}
\def\L{ {\cal L}}
\def\C{ {\cal C}}
\def\N{ {\cal N}}
\def\calE{ {\cal E}}
\def\calD{ {\cal D}}
\def\F{ {\cal F}}
\def\lin{{\rm lin}}
\def\Tr{{\rm Tr}}
\def\modS{{(S - \bar S)}}
\def\modT{{(T - \bar T)}}
\def\modU{{(U - \bar U)}}
\def\mpl{M_{\rm Pl}}

\newcommand{\Fg}[1]{{F}^{({#1})}}
\newcommand{\cFg}[1]{{\cal F}^{({#1})}}
\newcommand{\cFgc}[1]{{\cal F}^{({#1})\,{\rm cov}}}
\newcommand{\Fgc}[1]{{F}^{({#1})\,{\rm cov}}}
\def\mxth{\mathsurround=0pt }
\def\xversim#1#2{\lower2.pt\vbox{\baselineskip0pt \lineskip-.5pt
x  \ialign{$\mxth#1\hfil##\hfil$\crcr#2\crcr\sim\crcr}}}

\def\simgr{\mathrel{\mathpalette\xversim >}}
\def\simle{\mathrel{\mathpalette\xversim <}}

\newcommand{\ms}[1]{\mbox{\scriptsize #1}}
\renewcommand{\a}{\alpha}
\renewcommand{\b}{\beta}
\renewcommand{\c}{\gamma}
\renewcommand{\d}{\delta}
\newcommand{\pa}{\partial}
\newcommand{\g}{\gamma}
\newcommand{\G}{\Gamma}
\newcommand{\A}{\Alpha}
\newcommand{\B}{\Beta}
\newcommand{\D}{\Delta}
\newcommand{\e}{\epsilon}
\newcommand{\E}{\Epsilon}
\newcommand{\z}{\zeta}
\newcommand{\Z}{\Zeta}
\newcommand{\K}{\Kappa}
\renewcommand{\l}{\lambda}
\renewcommand{\L}{\Lambda}
\newcommand{\m}{\mu}
\newcommand{\M}{\Mu}
\newcommand{\n}{\nu}
\newcommand{\X}{\Chi}
\newcommand{\R}{\Rho}
\newcommand{\s}{\sigma}
\renewcommand{\S}{\Sigma}
\renewcommand{\t}{\tau}
\newcommand{\T}{\Tau}
\newcommand{\y}{\upsilon}
\newcommand{\Y}{\upsilon}
\renewcommand{\o}{\omega}
\newcommand{\q}{\theta}
\newcommand{\h}{\eta}

\def\dota{ {\dot{\alpha}} }
\def\lag{Lagrangian}
\def\Kahler{K\"{a}hler}
\def\A{ {\cal A}}
\def\C{ {\cal C}}
\def\F{{\cal F}}
\def\cL{ {\cal L}}

\def\R{ {\cal R}}
\def\x{ \times }
\def\ra{\rightarrow}
\def\la{\leftarrow}
\def\lra{\leftrightarrow}

\sloppy
\newcommand{\ov}{\overline}
\newcommand{\un}{\underline}
\newcommand{\p}{\partial}
\newcommand{\bm}{\boldmath}
\newcommand{\ds}{\displaystyle}
\newcommand{\nl}{\newline}
\newcommand{\Nzahl}{{\bf N}  }
\newcommand{\zzahl}{ {\bf Z} }
\newcommand{\Zzahl}{ {\bf Z} }
\newcommand{\Qzahl}{ {\bf Q}  }
\newcommand{\Rzahl}{ {\bf R} }
\newcommand{\Czahl}{ {\bf C} }
\newcommand{\wt}{\widetilde}
\newcommand{\wh}{\widehat}
\newcommand{\fs}[1]{\mbox{\scriptsize \bf #1}}
\newcommand{\ft}[1]{\mbox{\tiny \bf #1}}
\newtheorem{satz}{Satz}[section]
\newenvironment{Satz}{\begin{satz} \sf}{\end{satz}}
\newtheorem{definition}{Definition}[section]
\newenvironment{Definition}{\begin{definition} \rm}{\end{definition}}
\newtheorem{bem}{Bemerkung}
\newenvironment{Bem}{\begin{bem} \rm}{\end{bem}}
\newtheorem{bsp}{Beispiel}
\newenvironment{Bsp}{\begin{bsp} \rm}{\end{bsp}}
\renewcommand{\arraystretch}{1.5}


\renewcommand{\thesection}{\arabic{section}}
\renewcommand{\theequation}{\thesection.\arabic{equation}}

\parindent0em

\begin{titlepage}
\begin{center}
\hfill SLAC-PUB 7326\\
\hfill SNUTP-96/105\\
\hfill SU-ITP 96-44\\
\hfill {\tt hep-th/9610157}\\

\vskip .1in

{\bf 
CLASSICAL AND QUANTUM ASPECTS OF BPS BLACK HOLES \\
IN $N=2,D=4$ HETEROTIC STRING COMPACTIFICATIONS\footnote{
Work supported in part by DOE DE-AC03-76SF00515, NSF Grant PHY-9219345, 
NSF-KOSEF Bilateral Grant, KRF International Collaboration Grant and
Non-Directed Research Grant 81500-1341, KOSEF Purpose-Oriented Research
Grant 94-1400-04-01-3 and SRC Program, the Ministry of Education BSRI Grant
96-2418, and the Seoam Foundation Fellowship.}
}

\vskip .2in

{\bf Soo-Jong Rey}\footnote{email: 
\tt sjrey@leland.stanford.edu, sjrey@slac.stanford.edu, sjrey@phyb.snu.ac.kr}
\\
\vskip 0.5cm

{\em Stanford Linear Accelerator Center, Stanford University, 
     Stanford CA 94309 USA}
\\
{\em Department of Physics, Stanford University, Stanford CA 94305 USA}
\\
{\em Physics Department, Seoul National University, Seoul 151-742 KOREA}

\end{center}

\vskip .1in

\begin{center} {\bf ABSTRACT } \end{center}
\vskip-0.125in
\begin{quotation}\noindent
We study classical and quantum aspects of $D=4, N=2$ BPS black 
holes for $T_2$ compactification of $D=6, N=1$ heterotic string vacua. We 
extend {\sl dynamical relaxation phenomena} of moduli fields to background 
consisting of a BPS soliton or a black hole and provide a simpler but 
more general derivation of the Ferrara-Kallosh's extremized black hole 
mass and entropy. We study quantum effects to the the BPS black hole mass 
spectra and to their dynamical relaxation. We show that, despite 
non-renormalizability of string effective supergravity, quantum effect
modifies BPS mass spectra only through coupling constant and moduli field
renormalizations. Based on target-space duality, we establish a perturbative 
non-renormalization theorem and obtain exact BPS black hole 
mass and entropy in terms of renormalized string loop-counting parameter and 
renormalized moduli fields. We show that similar conclusion holds, in the 
large $T_2$ limit, for leading non-perturbative correction. We finally 
discuss implications to type-I and type-IIA Calabi-Yau black holes.
\end{quotation}
\vskip .3cm
\centerline{Submitted to Nuclear Physics B}
\end{titlepage}
\vfill
\eject

\newpage

\def\be{\begin{equation}}
\def\ee{\end{equation}}
\def\bee{\begin{eqnarray}}
\def\eee{\end{eqnarray}}

\section{INTRODUCTION}

\setcounter{equation}{0}
In recent exciting development~\cite{stromingervafa, horowitzstrominger,
maldacenacallan, myeretal}, string theory has provided with a microscopic
first--principle from which long--standing puzzle of black hole 
thermodynamics~\cite{hawking} can be understood. 
The development was made possible, on one hand, from better understanding of 
non-perturbative string theory including strong-weak coupling 
duality~\cite{ sduality}, various string-string dualities~\cite{hulltownsend, 
witten, kachruvafa, ferraravafa, polchinskiwitten, horavawitten} and 
D-brane soliton sector carrying Ramond-Ramond 
charges~\cite{polchinski}, and, on another, from 
deeper understanding of BPS states
in string theory and their stability throughout weak to strong coupling 
regime.

While a definite relation between the
statistical mechanics of microscopic stringy states and the macroscopic
laws of the black hole thermodynamics has been established for various
specific black holes, it has not yet provided a {\sl universal} 
derivation of entropy--area relation for {\sl all} 
classes of black holes. 
In particular, all specific examples explored so far have had large 
$N \ge 4$ supersymmetries in four dimensions. 
Together with the fact that the scalar fields at black hole horizon 
were fixed~\cite{wilczek,cvetictseytlin} so that the horizon area were 
independent of the scalar fields at infinity, large $N \ge 4$ 
local supersymmetries were 
stringent enough to determine the macroscopic 
black hole entropy uniquely (up to constants). 
In view of this, extension of the previous studies to the less stringent
yet controllable situations is posed as an interesting problem 
and might offer further insights to the black--hole physics. 
In this respect, black holes arising in $N=2$ supersymmetric theories 
are unique in that there exist controllable BPS states 
yet smaller supersymmetries renders 
underlying dynamics richer enough. 

Recently, initiated by the pioneering work of Ferrara, Kallosh and 
Strominger~\cite{ferrarakalloshstrominger}, macroscopic aspects of 
$N=2$ BPS black holes have been studied extensively
~\cite{strominger}--\cite{lustgroup}. In these works, the special
geometry~\cite{specialgeometry} that governs interactions of $N=2$ 
supergravity with vector and hyper multiplets has played an 
important role.  On the microscopic side, examples of D-brane 
configurations in $D=4, N=2$ compactifications have been 
found~\cite{strominger2} and microscopic state counting has shown  
complete agreement with macroscopic entropy formula of corresponding 
black holes.

Particularly interesting subset of $N=2$ BPS black holes are are the ones 
having constant moduli everywhere outside black hole 
horizon~\cite{strominger, ferrarakallosh}. These, so-called double
extreme black holes~\cite{kalloshstudent1}, are distinguished from other
BPS black holes in that they have the lightest possible mass.
With such special properties, one might 
expect that the double-extreme black holes play a special role among all
$N=2$ BPS black holes and open up new understanding uncovered so far. 
In this respect, better understanding of the double-extreme black holes
is desirable. The first motivation and contents of the present work is 
to address various aspects of them.

In all previous microscopic and macroscopic studies, however, only classical 
aspects of $N=2$ BPS black holes were considered. What distinguishes $N=2$ 
supersymmetry from $N \ge 4$ ones is that nontrivial quantum effects are 
present generically at perturbative and non-perturbative levels~\cite{
seibergwitten}.
In establishing entropy--area relation for black holes in $N \ge 4$ 
supersymmetric theories, stability of BPS states against strong coupling 
extrapolation has served an integral part of underlying physics.
Hence, it is of interest to what extent the nontrivial quantum effects are
reflected in the $N=2$ black holes and their physical properties.
The second motivation and contents of the present work is study 
quantum effects for BPS states in rigid and local $N=2$ theories.

In this paper, we study the above aspects for four--dimensional $N=2$ heterotic
string compactifications. For definiteness, we focus on rank-3, so-called
STU model, theories that arises from compactification on $T_2$ of a 
$D=6, N=1$ heterotic string theory, which have been obtained from $D=10$
by compactifying on $K_3$ with instanton numbers $(12,12)$\footnote{
This compactification is identical to other ones with different instanton
numbers $(10, 14)$ or $(11, 13)$ at least perturbatively. Non-perturbative
effects, however, may reveal differences among them\cite{berglund, theisen}
.}. This theory is known to be dual either to the type IIA compactification 
on the Calabi-Yau threefold ${\bf P}_{1,1,2,8,12}(24)$~\cite{kachruvafa}
 or to the $T_2$ compactification of the $D=6$ type-I 
orientifold on $K_3$ orbifold $T_4 / {\bf Z}_2$ with 
one tensor multiplet and completely Higgs gauge group~\cite{seiberggroup}.
Utilizing each duality map, we may learn otherwise inaccesible properties 
of type IIA and type I black holes from heterotic STU black holes as well. 

This paper is organized as follows. In Section 2, we first consider
rigid $N=2$ theory and interpret BPS mass minimization as dynamical 
relaxation of scalar fields. We extend this to local $N=2$ theory and
obtain what we call K\"ahler-BPS condition. This provides a simpler and
more general derivation of the result by Ferrara and 
Kallosh~\cite{ferrarakallosh}. In Section 3, we consider classical
aspect of K\"ahler-BPS black holes for the heterotic STU model. 
In Section 4, we study quantum effects to K\"ahler-BPS configurations. 
After recalling the rigid $N=2$ supersymmetry  non-renormalization theorems
to BPS masses, we study quantum effects to BPS black holes of the heterotic 
STU model. We show that target-space duality symmetry provides a strong 
constraint to quantum corrections. We establish a 
perturbative non-renormalization theorem based on the symmetry and
 show that the BPS black holes continues to saturate the BPS bound 
in terms of renormalized string loop-counting parameter and moduli fields. 
In the large 
$T_2$ limit, we also derive leading non-perturbative corrections.
In Section 5, we conclude with brief discussions on utilizing string-string 
duality maps to type--I and type--IIA string theory black holes.
\section{DYNAMICAL RELAXATION OF $N=2$ BPS MASS SPECTRA}
\setcounter{equation}{0}
\subsection{BPS Mass and Dynamical Relaxatioin in Rigid $N=2$ Theory}
\subsubsection{\rm RIGID $N=2$ SPECIAL GEOMETRY}
Before we dwell into the technically more involved $N=2$ supergravity theory, 
we first study dynamical relaxation of the extremal BPS mass spectra for
rigid $N=2$ supersymmetric Yang-Mills theory.
  The Lagrangian of the $N=2$ supersymmetric gauge theory is 
constructured out of $N=2$ vector multiplets.
The $N=2$ vector multiplets $X^A$ are constrained chiral superfields and 
contain a complex scalar, $SO(2)$ doublet Majorana-Weyl gauginos and an abelian 
gauge field. We denote them as $(X^A, \Omega^A_a, \Omega^{Aa}, 
A_\mu^A)$, $A = 1, \cdots, n_V; a = 1,2$ for both left-handed $\Omega_a$ and
right-handed $\Omega^a$'s. 
Associated with them, one introduces a holomorphic 
prepotential $F(X)$ of the vector multiplets. Coupling of the scalar fields 
with the vector field strengths is then specified by a holomorphic tensor:
\be
{\overline {\cal N}}_{AB} 
\equiv {\theta_{AB} \over 2 \pi} + i {4 \pi \over g^2_{AB}}
= {\partial}_{A} {\partial}_{B} 
{F}.
\label{couplingmatrix}
\ee
To describe scalar self-interaction, we first combine $X^A$ and 
$F_A \equiv \partial F / \partial X^A$ together, and construct a symplectic 
vector $V$:
\be
V = \left( \begin{array}{c} X^A \\ F_A \end{array} \right).
\label{sympvec}
\ee
Denoting symplectic inner product in terms of matrix multiplication
\be
\langle V | W \rangle \equiv V^t \cdot \omega \cdot W \,\,; 
\hskip1cm
\omega = \left( \begin{array}{cc} 0 & + {\bf I} \hskip-0.16cm {\bf I} \\ 
- {\bf I} \hskip-0.16cm {\bf I} & 0 \\ \end{array} \right),
\ee
K\"ahler potential defined by
\be
K(Z, {\overline Z}) = i \langle V | {\overline V} \rangle 
= i \Big( X^A {\overline F}_A ({\overline X})  - {\overline X}^A F_A(X) \Big).
\label{rigidkahler}
\ee
specifies a $n_V$-dimensional K\"ahler manifold. By adopting so-called 
`rigid special coordinates' $Z^A = X^A$, we find the K\"ahler metric
\be
K_{A{\overline B}} = \partial_A {\overline \partial}_B K 
= - 2 ({\rm Im} {\cal N})_{AB}.
\label{kahlermetric}
\ee
The fact that couplings of scalar self-interactions and scalar-vector
interactions are the same is nothing but a manifestation of the
underlying $N=2$
supersymmetry. By the same reason, the scalar-gaugino interaction
couplings should also be governed by ${\cal N}_{AB}$. Indeed, expanding
the Lagrangian defined by chiral $F$-term of the prepotential
into component fields, one obtains:
\be
L = - {1 \over 4 \pi} {\rm Im} \Big( 
{ 1 \over 2} {\overline {\cal N}}_{AB} {\cal F}^{-A}_{\mu \nu} 
{\cal F}^{-B \mu \nu} 
+ 2 {\overline {\cal N}}_{AB} {\cal D}_\mu X^A {\cal D}^\mu {\overline X}^B
+ {\overline {\cal N}}_{AB} {\overline \Omega}^{Aa}{\cal D} 
\hskip-0.23cm / \, \Omega^B_a \Big) + \cdots.  
\label{lag}
\ee
Here, the ellipses denote non-minimal coupling terms including magnetic
moment interactions and scalar potential.

Heuristically, one can draw an analogy to an electrodynamics
in a macroscopic media~\cite{jackson}. 
The coupling matrix ${\cal N}_{AB}$ is then 
interpreted as a generalized electric permittivity and magnetic permeability
tensors:
\bee
\epsilon_{AB} & = & [1 + \chi_E]_{AB} = [ \, {\rm Im} {\cal N} \, ]_{AB} 
\nonumber \\
\mu^{AB} &=& [1 + \chi_M]^{AB} = [({\rm Im} {\cal N})^{-1}]^{AB}
\eee
The novelty of the analog macroscopic media is that the speed of light 
remains unity as can be confirmed from the fact 
$\epsilon_{AB} \mu^{BC} = 
[{\rm Im} {\cal N}]_{AB} [({\rm Im} {\cal N})^{-1}]^{BC} = \delta_A^C$.
This is a necessary condition to maintain the manifest Lorentz covariance
of the theory\footnote{ We note that effective supergravity theory of 
the type-I open string provides an another interesting analog macroscopic 
media with a unit speed of light, 
for which the electric permittivity and the magnetic permeability
is self-field dependent:
\be
[\epsilon]^\mu_\nu = [{1 \over \mu}]^\mu_\nu 
= \Big[ 1 - {1 \over 4 \pi \alpha'}
F^{\alpha \beta} F_{\alpha \beta} \Big]^{-{1/2}} \delta^\mu_\nu.
\ee
}
The analogy with macroscopic media turns out quite useful later for 
interpreting dynamical relaxation of the BPS mass as a result of 
(anti)-screening of microscopic electric and magnetic charges by 
the macroscopic media.
Motivated by this analogy, one then introduce a symplectic vector of 
the anti-self-dual field strengths:
\be
{\cal Z}^- = ({\cal F}^{-A}, {\cal G}^-_A).
\label{asdfieldst}
\ee
Symplectic vector of self-dual field strengths ${\cal Z}^+$ is 
defined by a complex conjugate relation of Eq.(~\ref{asdfieldst}).
The field ${\cal F}^A$ corresponds to the generalized electric and magnetic
induction fields, ${\bf E}$ and ${\bf B}$. Similarly, the field ${\cal G}_A$
corresponds to the generalized electric displacement and magnetic fields,
${\bf D}$ and ${\bf H}$.   
These two sets of field-strength sections are related each other 
\be
{\cal G}^-_A \equiv {\overline {\cal N}}_{AB} {\cal F}^{-B} .
\label{constitutive}
\ee
This is a direct counterpart of the so-called `constitutive relation'~\cite{
jackson} in the electrodynamics of a macroscopic media, viz., 
a functioinal relation
of ${\bf D}, {\bf H}$ in terms of ${\bf E}, {\bf B}$.

In the presence of electric and magnetic four-currents
$ {\cal J} = (J^A_e, J_{Am} )$, $8 n_V$-component Maxwell's equation 
is expressed compactly as:
\be
{\rm d} \wedge {\rm Re} {\cal Z}^- = \wedge {\cal J}.
\ee
Integrating this equation over the space, we obtain a symplectic
vector of microscopic electric and magnetic charges:
\bee
{\cal Q} &\equiv& (P^A, Q_A), \nonumber \\
\oint_{S_2} {\cal F}^A &=& P^A \,\, ; \hskip1cm \oint_{S_2} {\cal G}_A
= Q_A.
\label{Qcharge}
\eee Classically, the charges are continuous, real-valued in units of
appropriate electric and magnetic coupling constants.  Quantum
mechanically, however, the charges should obey the
Dirac-Schwinger-Zwanziger quantization condition, $ \langle {\cal Q} |
{\cal Q}' \rangle = P^\Lambda Q_\Lambda' - Q_\Lambda P^{'\Lambda} $ 
is an integer multiple of the Dirac unit $(2 \pi \hbar)$.
This in turn implies that the symplectic charge vector
${\cal Q}$ is covariant only under $Sp(2n_V; {\bf Z}) \in Sp(2n_V;
{\bf R})$.

\subsubsection{\rm $N=2$ CENTRAL CHARGE AND BPS SPECTRA}
One can derive the central charge directly from the supersymmetry algebra.
The supersymmetry current $S_{\mu a}$ and the supercharge of Eq.(\ref{lag}) 
are given by:
\bee
S_\mu^a &=& - ({\rm Im} {\cal N})_{AB} 
\Big[\gamma_\nu \gamma_\mu 
\Omega^A_a {\cal D}^\nu {\overline X}^B - \epsilon_{ab} 
\gamma^\alpha \gamma^\beta \gamma_\mu \Omega^{Ab} {\cal F}^{-B}_{\alpha \beta}
\Big];
\nonumber \\
{\bf Q}_a &\equiv& \int d^3 {\bf x} \, S^0_a.
\eee
From Eq.(\ref{lag}), one also derives anti-commutation relations 
for the gaugino fields:
\be
\{ \Omega^A_a (t, {\bf x}), \Omega^{\dagger B b} (t, {\bf y}) \}_{\rm ET} 
=
\{ \Omega^{Aa} (t, {\bf x}), \Omega^{\dagger B}_b (t, {\bf y}) \}_{\rm ET}
= \Big[{ i \over {\rm Im} {\cal N}} \Big]^{AB} \delta^a_b 
\delta ( {\bf x} - {\bf y} ).
\ee
One then evaluates the supercharge anti-commutators:
\bee
\{ {\bf Q}_a, {\overline {\bf Q}}^b \}
&=& + \delta_a^b \gamma^\mu P_\mu, 
\nonumber \\
\{{\bf Q}_a, {\overline {\bf Q}}_b \}
&=& - \epsilon_{ab} \int d^3 {\bf x} 
\Big[ {\vec {\cal G}}_A \cdot {\vec {\cal D}} X^A -  
{\vec {\cal F}}^A \cdot {\vec {\cal D}} {\overline F}_A \Big].
\eee
The central charge is defined from the second anti-commutator after 
integrating by parts and using the Maxwell's equation Eq.(\ref{Qcharge}):
\bee
{\bf Z} &\equiv& X^A \oint {\cal G}_A - F_A \oint {\cal F}^A
= X^A Q_A - F_A P^A \nonumber \\
&=& \langle V | {\cal Q} \rangle.
\label{centralch}
\eee
Note the topological nature of the central charge as it is defined as the
surface integral at spatial infinity. Diagonalizing the supersymmetry 
algebra, one finds the BPS inequality for the mass spectra:
\be
{\bf M}^2 \ge {\bf M}^2_{\rm BPS} \equiv
|{\bf Z}|^2 = | \langle V | {\cal Q} \rangle |^2.
\ee

\subsubsection{\rm DYNAMICAL RELAXATION OF BPS MASS}
We now introduce a notion of {\sl dynamical relaxation of BPS mass
and free energy}.
Consider a single, isolated BPS state carrying electric and magnetic
charges specified by the symplectic vector ${\cal Q}$.
The BPS mass
${\bf M}_{\rm BPS}$ defines a mass gap separating the BPS state from vacuum
state, and is a function of the gauge coupling constants.
In supersymmetric gauge theory, the gauge coupling matrix ${\cal N}_{AB}$
is not a constant but a function of the coordinates $Z^A$ on K\"ahler manifold.
These coordinates are dynamical fields in $N=2$ supersymmetric
gauge theory, hence, can relax dynamically and minimize the mass gap
${\bf M}_{\rm BPS}$.
Since the BPS state is characterized by electric and magnetic charges it
carries, the relaxation configuration of the special coordinates
$Z^A$ should be determined entirely in terms of the symplectic charge vector
${\cal Q}$.
Therefore, up to $Sp(2n_V; {\bf R})$ symplectic transformations,
one can associate a one-to-one mapping
from the $n_V$-dimensional  K\"ahler manifold parametrized
by the scalar fields $X^A = Z^A$ to the space of electric and magnetic charges
of a given BPS state. Such a mapping is a harmonic one and the BPS mass
can be taken as the positive-definite free energy associated with the
harmonic mapping. Obviously, this notion can be extended to situations
of multiple BPS states.
  
To exemplify the notion of dynamical minimization of BPS mass gap,
consider a situation for the gauge group of rank one.
The BPS mass spectra may be written as
\be
{\bf M}^2_{BPS} = {\tt M}_{\rm W}^2 \, [Q^2 +  ({ 4 \pi \over g^2} \, P)^2 ].
\ee
Here, ${\tt M}_{\rm W}$ denotes mass of heavy charged gauge boson and
is related to vacuum expectation value $v$ of Higgs fields as
${\tt M}_{\rm W} = g v$. 
Using the Cauchy-Schwarz inequality~\cite{inequality},
\be
{\bf M}^2_{\rm BPS} \ge 2 {{\tt M}_{\rm W}^2 \ \alpha_g} \, 
\sqrt { P^2 Q^2 }  = { 4 \pi \over g^2} \, [ 8 \pi v^2] \, P^2,
\ee
and the equality is saturated at $ \alpha_g \equiv 
{g^2 \over 4 \pi}  = \sqrt {P^2 / Q^2}$.

It is straightforward to generalize the example to rank-$N$ ($N > 1$) 
gauge group. For simplicity, we consider the minimal coupling 
${\overline {\cal N}}_{AB} = 
[(\theta/2 \pi) + i (4 \pi / g^2)] \delta_{AB}$, but the foregoing
result can be generalized to non-minimal case straightforwardly .
The gauge group is spontaneously
broken to $[U(1)]^N$. Taking into account of the Witten 
effect~\cite{witteneffect} and introducing a
notation $\langle A, B \rangle $ for the quadratic form in the charge
space, the BPS mass spectra is given by 
\be
{\bf M}^2_{\rm BPS} = {\tt M}_{\rm W}^2 \, [ \langle 
( Q - {\theta \over 2 \pi}  P ) ,
( Q - {\theta \over 2 \pi} P ) \rangle
+  ({ 4 \pi \over g^2})^2 \langle P, P \rangle ].
\label{newcharge}
\ee
By introducing a new basis of electric and magnetic charges:
\be
{\bf p} \equiv P; \hskip1cm {\bf q} \equiv {g^2 \over 4 \pi} (Q - {\theta 
\over 2 \pi} P),
\ee
one has
\be
{\bf M}^2_{\rm BPS} 
= {\tt M}_{\rm W}^2 \, \Big({4 \pi \over g^2} \Big)^2 \,
\Big[ \langle {\bf p}, {\bf p} \rangle 
+ \langle {\bf q}, {\bf q} \rangle \Big].
\label{newbps}
\ee  
Again, one finds that there exists a special configuration of the 
coupling constants at which BPS mass gap is minimized:
\bee
{\bf M}^2_{\rm BPS} &\ge&  {{\tt M}_{\rm W}^2 \over \alpha^2_g} \, 
\sqrt{ \langle {\bf p}, {\bf p} \rangle \,
\langle {\bf q}, {\bf q} \rangle}
\nonumber \\
&\ge& {{\tt M}_{\rm W}^2 \over \alpha_g^2} \, 
( {\bf p} \cdot {\bf q} ) \nonumber \\
&\ge & 0.
\label{bpsineq}
\eee
Here, we have used the Cauchy-Schwarz inequality 
and the H\"older's inequalities~\cite{inequality} at the first and 
the second steps respectively. Each inequality then provides 
with separate conditions to the coupling constant $g^2 / 4 \pi$ and 
to the vacuum angle $\theta / 2 \pi$ for which the BPS mass is minimized. 
First, the extremal vacuum angle is determined by saturating 
the H\"older's inequality, viz., the second line in Eq.(\ref{bpsineq}):
\bee
({\bf p} \cdot {\bf q} ) &=& \alpha_g \, \langle P, ( Q - {\theta \over 
2 \pi} P ) \rangle = 0;
\nonumber \\
\rightarrow \hskip0.75cm {\theta \over 2 \pi} &=& { ( P  \cdot Q ) \over
\langle P ,  P \rangle }.
\label{extangle}
\eee
Next, the extremal coupliing constant is determined by saturating the
Cauchy-Schwarz inequality, viz., the first line in Eq.(\ref{bpsineq}):
\bee
\langle P, P \rangle &=& \langle {\bf p}, {\bf p} \rangle 
= \langle {\bf q}, {\bf q} \rangle =
\Big( { g^2 \over 4 \pi} \big)^2 \, \langle ( 
Q - {\theta \over 4 \pi} P ), (Q - {\theta \over 4 \pi} P) \rangle;
\nonumber \\
\rightarrow \hskip0.75cm 
{ 4 \pi \over g^2} &=& { 1 \over \langle P, P \rangle}
\sqrt { \langle Q , Q \rangle \langle P, P \rangle
- ( P \cdot Q )^2 }.
\label{extcoupl1}
\eee
To obtain the last expression, we have inserted the extremal vacuum angle
Eq.(\ref{extangle}).
Alternatively, one may first tune the the vacuum angle to $\theta = 0$
by a Peccei-Quinn transformation and determine the extremal gauge coupling 
constant by Cauchy-Schwarz inequality. It is given by:
\be
[{4 \pi \over g^2}]_{\theta = 0} = \sqrt {\langle Q , Q \rangle
\over \langle P , P \rangle }.
\ee
One then undo the Peccei-Quinn rotation of the vacuum angle $\theta$. 
Recalling that the vacuum angle shift also 
introduces an electric charge by the Witten effect~\cite{witteneffect},
one finds that the BPS mass minimized under the condition $\theta = 0$ 
can be lowered further.  Saturation  of the new BPS mass then yields 
exactly the same result as the one based on the H\"older's inequality in
Eq.(~\ref{bpsineq}). Hence, the
vacuum angle relaxes to the extremal value Eq.(~\ref{extangle}) and, in
turn, the coupling constant further to the extremal value Eq.(~\ref{extcoupl1}).
In either methods, one finally obtain the extremal BPS mass gap or free energy:
\be
{\bf M}^2_{\rm BPS} = {{\tt M}_{\rm W}^2 \over \alpha_g} \, 
\sqrt { \langle Q, Q \rangle
\langle P , P \rangle
- ( P \cdot Q)^2 }
= { 4 \pi  \over g^2 } \, [ 8 \pi v^2] \, 
\langle P , P \rangle.
  \label{extcoupl2}
\ee

The main idea of the BPS mass minimization is that
the the gauge coupling constants parametrized by $n_V$ complex scalar fields
on the K\"ahler manifold are actually not constants but can relax. 
Since the BPS mass-squared is a positive-definite quadratic form of 
$2n_V$ electric and magnetic charges, it can be taken as a free energy 
that determines the relaxation configuration. 
In the analog electrodynamics of a macroscopic media, one allows the
electric permittivity and the magnetic permeability to relax dynamically
so that the screening of electric and magnetic monopole charges becomes
as perfect as possible. Because of the Lorentz covariance relation
$\epsilon \cdot \mu = 1$, screening of electric and screening of magnetic
charges compete each other. The above extremal configuration is where 
the competition is balanced.        
While it is rather artificial in non-supersymmetric theories, the notion 
of dynamical relaxation is quite natural in supersymmetric field theories, 
supergravity theories and superstring theory. In fact, the idea 
has been used repeatedly for minimizing vacuum energy and determine 
physical parameters dynamically for various 
situations~\cite{dynamicalrelax}. The only novelty in the present situation is 
that the background under consideration is not a flat spacetime but 
a BPS soliton or, as we will extend later, a black hole carrying 
nonvanishing electric and magnetic charges.

\subsection{Local Special Geometry and $N=2$ Supergravity}
\subsubsection{\rm LOCAL $N=2$ SPECIAL GEOMETRY}
Consider the space of the $n_V$ complex scalar fields $Z^A$ associated
with vector multiplets in the $N=2$ supergravity. Locally this space 
form a K\"ahler-Hodge manifold, endowed with a K\"ahler potential 
$K(Z, {\overline Z})$ and a K\"ahler metric $K_{A {\overline B}}
\equiv \partial_A {\overline \partial}_B K$.
The local $N=2$ supergravity algebra constrains that the Riemann 
curvature tensor of the K\"ahler-Hodge manifold should obey so-called
`special-geometry' relation:
\be
R_{A {\overline B} C {\overline D}}
= K_{A {\overline B}} K_{C {\overline D}} + 
K_{A {\overline D}} K_{C {\overline B}} 
- Y_{ACE} {\overline Y}_{\overline BDF} K^{E {\overline F}}.
\label{spegeorel}
\ee

To define the $N=2$ supergravity couplings, 
we start by defining symplectic sections of the Hodge bundle: 
\be
V \equiv (L^\Lambda, M_\Lambda) ; \hskip1cm \Lambda = 0, 1, \cdots, n_V.
\ee
They are covariantly holomorphic
\be
D_{\overline A} V \equiv [ {\overline \partial}_{\overline A} - { 1 \over 2} 
K_{\overline A} ] V = 0.
\ee
Projection of $L^\Lambda$'s to $Z^A$'s is achieved by a gauge fixing. 
Demanding that the scalar and the graviton kinetic terms decouple, we find 
the choice:
\be
\langle V | {\overline V} \rangle = (L^\Lambda {\overline M}_\Lambda
- {\overline L}^\Lambda M_\Lambda) = i.
\label{gf}
\ee
In addition to the section $V$, one can construct $n_V$ new 
symplectic sections $U_A$ out of $V$:
\be
U_A = D_A V \equiv (\partial_A + {1 \over 2} \partial_A K) V.
\ee
Then the special-geometry constraint Eq.(~\ref{spegeorel}) 
is solved by the above $(n_V + 1)$ symplectic sections
if they satisfy $(n_V+1)$ relations:
\be
V = \left( \begin{array}{cc} 0 & {\cal N}^{-1} \\ {\cal N} & 0 \\ \end{array}
\right) \cdot V; \hskip0.76cm 
U_A = \left( \begin{array}{cc} 0 & {\overline {\cal N}}^{-1} \\
{\overline {\cal N}} & 0 \\ \end{array} \right) U_A.
\ee
Here, a symmetric matrix ${\cal N}$ is solved by combining the 
$V, {\overline U}_A$ sections together into $(n_V + 1) \times (n_V + 1)$ 
matrix $(V, {\overline U}_{\overline A})^T$ and inverting the $(n_V + 1)$ 
relations:
\be
\left( \begin{array}{c} V \\ {\overline U}_{\overline A} \\ \end{array} 
\right) 
= \left( \begin{array}{cc}
0 & + {\cal N}^{-1} \\ {\cal N} & 0 \\ \end{array} \right) 
\left( \begin{array}{c}  V \\ {\overline U}_{\overline A} \\ \end{array} 
\right)
 \hskip0.5cm \rightarrow \hskip0.5cm 
\left( \begin{array}{cc} 0 & {\cal N}^{-1} \\ {\cal N}&0\\ \end{array} \right)
 = \left( \begin{array}{c} V \\ {\overline U}_{\overline A} \\ \end{array} 
\right)
\cdot 
\left( \begin{array}{c} V \\ {\overline U}_{\overline A} \\ \end{array} 
\right)^{-1}. 
\ee
Note that the gauge
fixing condition Eq.(~\ref{gf}) becomes 
\be
({\cal N}_{\Lambda \Sigma} - {\overline {\cal N}}_{\Lambda \Sigma})
L^\Lambda {\overline L}^\Sigma = i.
\label{newgf}
\ee
It is straightforward to solve ${\cal N}$ in terms of the symplectic 
sections and holomorphic matrix ${\cal F}$:
\be
{\cal N}_{\Lambda \Sigma} 
= {\overline {\cal F}}_{\Lambda \Sigma} 
+ 2i 
{ 
( {\rm Im} {\cal F})_{\Lambda \Gamma} L^\Gamma 
({\rm Im} {\cal F})_{\Sigma \Delta} L^\Delta \over
L^\Gamma ({\rm Im} {\cal F})_{\Gamma \Delta} L^\Delta };
\hskip0.5cm {\cal F}_{\Lambda \Sigma} \equiv
\partial_\Lambda F_\Sigma (X).
\ee
One further finds that 
\bee
\langle V | U_A \rangle &=& \langle V | {\overline U}_{\overline A} 
\rangle = 0, \nonumber \\
 K_{A {\overline B}} &=& - i \langle U_A | U_{\overline B} \rangle
, \nonumber \\
 Y_{ABC} &=& \langle D_A U_B | U_C \rangle.
\label{ortho}
\eee

It is possible to fix an overall scale of the symplectic section as
\be
V = {\rm M}_{\rm Pl} e^{K/2} \Omega; \hskip0.75cm \Omega \equiv
\left( \begin{array}{c} X^\Lambda \\ F_\Lambda \end{array} \right).
\ee
It follows that $\Omega$ are holomorphic sections over a line bundle.
All the above relations are then straightforwardly rewritten in terms of
the holomorphic sections. In terms of $\Omega$, the K\"ahler potential
is given by
\be
K = - \log [ i \langle \Omega | {\overline \Omega} \rangle ].
\ee
Under the K\"ahler transformation $K \rightarrow K + \Lambda + {\overline
\Lambda}$, $\Omega \rightarrow e^{-\Lambda} \Omega$. Therefore $X^\Lambda$
provides a homogeneous local coordinate system on the K\"ahler manifold.
One possible choice of the coordinate system is so-called `special 
coordinates':
\be
Z^A = {X^A \over X^0}.
\ee

The electric and the magnetic charges provide with the source to the 
black hole mass. To manifest the $Sp(2n_V + 2)$ symplectic structure,
it is convenient to introduce anti-self-dual field strengths:
\be
{\cal Z}^- \equiv ({\cal F}^{-\Lambda}, {\cal G}^-_\Lambda);
\hskip1cm {\cal G}^-_\Lambda = {\overline {\cal N}}_{\Lambda 
\Sigma} {\cal F}^{-\Sigma}.
\ee
The equations of motion and the Bianchi identities are then compactly expressed
as 
\be
{\rm d} \wedge ({\rm Re} {\cal Z}^- ) = {} \wedge {\cal J}.
\label{vectorfields}
\ee
As in the rigid theory,
the electric and the magnetic charges combine to a symplectic vector:
\be
{\cal Q} = (P^\Lambda, Q_{\Lambda});
\hskip1.5cm 
\oint_{S_2} {\rm Re} {\cal F}^{-\Lambda} = P^\Lambda,
\hskip0.75cm 
\oint_{S_2} {\rm Re} {\cal G}^-_\Lambda = Q_\Lambda.
\label{chargevector}
\ee
Again, quantum mechanically, the electric and the magnetic charges are 
required to obey the Dirac-Schwinger-Zwanziger quantization conditions.
Therefore, the charge symplectic vector ${\cal Q}$ is covariant only
under the $Sp(2n_V + 2; {\bf Z})$ transformations.

\subsubsection{\rm CENTRAL CHARGE AND BPS MASS SPECTRA}
For a manifest supersymmetric multiplet formulation, it turns out convenient
to reorganize the gauge field strengths into a new $(n_V + 1)$ linearly 
independent combinations:
\bee
T^- &\equiv& \,\, \langle V | {\cal Z}^- \rangle \,\, = (M_\Lambda {\cal F}^{-
\Lambda} - L^\Lambda {\cal G}^-_{\Lambda} ),
\nonumber \\
{ F}^{-A} &\equiv& G^{A {\overline B}} \langle 
{\overline U}_{\overline B} | {\cal Z}^- \rangle =
G^{A {\overline B}} ({\overline D}_B {\overline M}_\Lambda {\cal F}^{-\Lambda}
- {\overline D}_B {\overline L}^\Lambda {\cal G}_\Lambda^- ).
\label{newfieldst}
\eee
That there are no other linearly independent field strength combinations 
is easy to understand from the two identities:
\be
\langle {\overline V} | {\cal Z}^- \rangle = 0 = 
\langle U_A | {\cal Z}^- \rangle.
\ee
The $T^-$ and $F^-_A, \, (A = 1, \cdots, n_V)$
are the gravi-photon of the supergravity multiplet and the gauge fields
of $n_V$ vector multiplets respectively.

Associated to the new $(n_V + 1)$ linearly independent combinations of the
gauge field strengths are complex-valued, $(n_V + 1)$-component central 
charge vector:
\bee
{\bf Z} &\equiv & -{1 \over 2} \oint_{S_2} T^- = \langle V | {\cal Q} 
\rangle = 
(L^\Lambda Q_\Lambda - M_\Lambda P^\Lambda),
\label{centralcharge1} \\
{\bf Z}_A &\equiv&  -{1 \over 2} \oint_{S_2} G_{A{\overline B}} 
F^{+{\overline B}} = ( Q_\Lambda D_A L^\Lambda  -
P^\Lambda D_A M_\Lambda) = \langle U_A | {\cal Q} \rangle =
 D_A {\bf Z}.
\label{centralcharge2}
\eee
One notes that, under the K\"ahler transformation 
$K \rightarrow K + \Lambda + {\overline \Lambda}$,
the central charge transforms as holomorphic sections:
${\bf Z} \rightarrow e^{-\Lambda} {\bf Z}$, $ {\bf Z}_A
\rightarrow e^{-\Lambda} {\bf Z}_A$ .
These central charge vectors satisfy quadratic sum rules
\bee
|{\bf Z}|^2 + |{\bf Z}_A|^2 
&=& -{1 \over 2} {\cal Q}^T \cdot {\bf M}({\cal N}) \cdot {\cal Q},
\nonumber \\
|{\bf Z}|^2 - |{\bf Z}_A|^2 &=& 
- {1 \over 2} {\cal Q}^T \cdot {\bf M}({\cal F}) \cdot {\cal Q}.
\label{quadsums}
\eee
where
\be
{\bf M}({\cal N}) = {\cal R}^T ({\rm Re} {\cal N}) \cdot
{\cal D} ({\rm Im} {\cal N}) \cdot {\cal R} ({\rm Re} {\cal N})
\label{quad}
\ee
and
\be
{\cal R} ({\rm Re} {\cal N}) = \left( \begin{array} {cc} 
{\bf I} \hskip-0.16cm {\bf I}  & 0 \\
- {\rm Re} {\cal N} & {\bf I} \hskip-0.16cm {\bf I} \\ \end{array} \right);
\hskip1cm
{\cal D} ({\rm Im} {\cal N}) = \left( \begin{array} {cc}
{\rm Im} {\cal N} & 0 \\
0 & ({\rm Im} {\cal N})^{-1} \\ \end{array} \right).
\label{rdmatrix}
\ee
The ${\cal R} ({\rm Re} {\cal N})$ matrix defines a
symplectic transformation associated with the Witten 
effect~\cite{witteneffect}:
\bee
{\cal R}^T \left( \begin{array}{cc} 
0 & + {\bf I} \hskip-0.16cm {\bf I} \\ 
- {\bf I} \hskip-0.16cm {\bf I} & 0 \\ \end{array} \right)
{\cal R} &=& \left( \begin{array}{cc}  
0 & + {\bf I} \hskip-0.16cm {\bf I} \\
- {\bf I} \hskip-0.16cm {\bf I} & 0 \\ \end{array} \right);
\nonumber \\
{\cal R} \left( \begin{array}{cc}
0 & + {\bf I} \hskip-0.16cm {\bf I} \\
- {\bf I} \hskip-0.16cm {\bf I} & 0 \\ \end{array} \right)
{\cal R} &=& \left( \begin{array}{cc}
0 & + {\bf I} \hskip-0.16cm {\bf I} \\
- {\bf I} \hskip-0.16cm {\bf I} & 0 \\ \end{array} \right).
\label{rsymp}
\eee
Similarly, the matrices ${\bf M}({\cal F})$, ${\cal R} ({\rm Re} {\cal F})$
and ${\cal D} ({\rm Im} {\cal F})$ are defined by replacing ${\cal N}$ in
Eqs.(~\ref{quad},~\ref{rdmatrix}) into ${\cal F}$ respectively.
In terms of factorized matrices, the quadratic sum rules Eq.(\ref{quadsums})
are given by:
\bee
|{\bf Z}|^2 
+ |{\bf Z}_A|^2 &=& -{1 \over 2} 
(Q - {\overline {\cal N}} \cdot P)_\Lambda 
\Big[{1 \over {\rm Im} {\cal N}} \Big]^{\Lambda \Sigma}
(Q - {\cal N} \cdot P)_\Sigma;
\nonumber \\
|{\bf Z}|^2 - |{\bf Z}_A|^2 &=&
-{1 \over 2}
(Q - {\overline {\cal F}} \cdot P )_\Lambda
\Big[ \, {1 \over {\rm Im} {\cal F} }  \, \Big]^{\Lambda \Sigma}
(Q - {\cal F} \cdot P)_\Sigma.
\label{quadsumstwo}
\eee
The two formula shows clearly that ${\rm Im} {\cal N}$ defines 
a negative-definite metric, while ${\rm Im} {\cal F}$ defines a metric
of signature $(1, n_V)$ in the quadratic central charge sum rules.

Since the central charge and its derivatives are projections of the electric
and the magnetic charges with respect to the symplectic sections, they are
in general functions of moduli fields and complex-valued. The utility of the
new 
linear combinations of the gauge field strengths and associated 
the central charges becomes transparent once one solves the condition for
nontrivial BPS black holes to exist. We now turn to these conditions.

\subsection{$N=2$ Supersymmetric Black Holes}
Consider $N=2$ supergravity theory coupled to $n_V$ vector multiplets.
Explicit construction of the $N=2$ supersymmetric black hole utilizing
the special geometry was initiated by Ferrara, Kallosh and Strominger  
~\cite{ferrarakalloshstrominger}.

One obtains the metric, the $(n_V + 1)$ gauge fields including the 
gravi-photon field and the $n_V$ scalar fields $Z^A$ configurations
by solving the supersymmetry Killing spinor conditions
to the gravitino and the gaugino supersymmetry transformation rules:
\bee
\delta \Psi_{\mu a} &=& {\cal D}_\mu \epsilon_a + 
T^-_{\mu \nu} \gamma^\nu \, \epsilon_{ab} \epsilon^b,  \nonumber \\
\delta \Omega^{Aa} &=& i (\nabla \hskip-0.26cm / Z^A) \epsilon^a
+ F^{-A}_{\mu \nu} \gamma^\mu \gamma^\nu \, \epsilon_b.
\label{susytransf}
\eee
The classical, supersymmetric black hole configuration 
is obtained by demanding an existence of covariantly constant
spinors, 
$\delta_\epsilon \psi_{\mu}^a = \delta_\epsilon \Omega^\Lambda_a = 0$
with an ansatz
\bee
ds^2 &=& e^{+2U} dt^2 - e^{-2U} d {\vec x}^2 \,\, ; \hskip0.75cm
e^{-U} = (1 + {M \over r}) ,
\nonumber \\
{\cal F}^\Lambda &=& {{\bf q}^\Lambda \over r^2} \, [e^{2U} dt \wedge dr] 
+ {{\bf p}^\Lambda \over r^2} \, [r^2 d \Omega_2].
\label{bhansatz}
\eee
Here, ${\bf p}$ and ${\bf q}$ are arbitrary constant denoting the
magnetic and electric charges measured from $F^\Lambda$ field at 
spatial infinity.
Recall that, in an analogy with the electrodynamics of a macroscopic media,
${\cal F}^\Lambda$ field corresponds to the generalized electric and magnetic
induction fields, ${\bf E}$ and ${\bf B}$. Therefore, one expects that the
electric charge ${\bf q}$ is not the microscopic charge but the total
charge including the screening and the Witten effect~\cite{witteneffect}.
Given the constitutive relations Eq.(\ref{vectorfields}), one should then
find a relation to the fundamental, microscopic charges ${\cal Q} 
= (P, Q)$. By a straightforward calculation, one finds that
\bee
Q_\Lambda &=& ({\rm Re} {\cal N})_{\Lambda \Sigma} {\bf p}^\Sigma 
- ({\rm Im} {\cal N})_{\Lambda \Sigma} {\bf q}^\Sigma;
\nonumber \\
P^\Lambda &=& {\bf p}^\Lambda.
\eee
The afore-mentioned screening and Witten effects are manifest from the 
charge relations.

A particularly interesting class of the $N=2$ black hole configurations
is the ones with a frozen special coordinate fields, $Z^A = $ 
constant~\cite{strominger, ferrarakallosh}.
The gaugino supersymmetry transformation rules in Eq.(\ref{susytransf}) 
then imply that the
gauge fields associated with the vector multiplets should vanish 
everywhere:
\be
F^{-A} = 0 \hskip0.75cm \leftrightarrow \hskip0.75cm 
F^{+ {\overline A}} = 0.
\ee
Inferring Eq.(\ref{centralcharge2}) for the above gauge field 
configuration
, one finds
\be
{\bf Z}_A = D_A {\bf Z} = 0,
\label{mincond}
\ee
and concludes that the black holes with frozen special coordinate 
fields exhibit 
a special feature that central charge ${\bf Z}$ is covariantly 
constant with respect to the special coordinates. 
Being $n_V$ independent equations, Eq.(\ref{mincond}) in turn 
determines uniquely the configuration of the $n_V$ special coordinates
$Z^A$.

The fact that the constant moduli ansatz leads to the lowest BPS 
mass of the black hole may be understood as follows. The total 
energy ${\cal E}$ of the black hole configuration may be expressed 
schematically as:
\be
{\cal E}_{\rm BH} 
= \int_{M_3}
\Big[ R^{(3)} + || \nabla Z^A ||^2 + 
{1 \over 2} \Big( {\bf E} \cdot {\bf D} (Z) + {\bf B} \cdot {\bf H}(Z) 
\Big) \Big].
\ee
That this expression is positive-definite is guaranteed by the Witten's
positive energy theorem~\cite{wittenposenergy},
 applied to the background of a black hole with nonvanishing electric 
and magnetic charges~\cite{gibbonshull}.
The second term represents a harmonic map to the K\"ahler manifold
whose coordinates are represented by the complex scalar fields $Z^A$.
Since the weight $|| \nabla  Z^A||^2$ is manifestly positive definite,
the lowest but nonzero BPS energy is achieved by a {\sl constant } 
harmonic map: $\nabla Z^A = 0$, viz. map the entire
space of the black hole exterior (outside the horizon) to a single
point in the K\"ahler manifold. 
Because of $N=2$ supersymmetry, this in turn requires that the 
the gauge field strengths $F^A$ paired with $Z^A$ fields vanish 
as well.
In what follows, we will call the constant harmonic map as 
{\sl K\"ahler-BPS limit}, since it follows from the minimization 
of the K\"ahler sigma model contribution to the BPS black hole energy. 
\subsection{Dynamical Relaxation of the BPS Black Hole Spectra}
\subsubsection{\rm K\"AHLER-BPS SCALAR FIELDS}
One now solves the K\"ahler-BPS condition and determines explicitly
the constant scalar fields $Z^A$ as a function of charges:
\be
{\bf Z}_A \equiv D_A {\bf Z} = 0 \hskip1cm 
\leftrightarrow \hskip1cm \langle  U_A | {\cal Q} \rangle = 0.
\label{mincond2}
\ee
In this case the two quadratic chrage 
sum rules Eq.(\ref{quadsumstwo}) reduce down to the square of the
central charge ${\bf Z}$ itself:
\bee
|{\bf Z}|^2 &=& -{1 \over 2} {\cal Q}^T \cdot {\bf M} ({\cal N}) \cdot {\cal Q}
\nonumber \\
&=& - {1 \over 2} {\cal Q}^T \cdot {\bf M}({\cal F}) \cdot {\cal Q}.
\label{extcentralcharge}
\eee 
While ${\cal N}$ and ${\cal F}$ matrices are different in general, for the
sum rules of the minimized central charge, quadratic forms formed out of
either matrices are the same. In subsequent calculations, we will
use exclusively the quadratic form using the holomorphic matrix
${\cal F}$ mainly for calculational convenience. However, all the formulas
we derive in this paper are straightforwardly generalizable to the
representation using the coupling
matrix ${\cal N}$  by replacing  wherever ${\cal F}$ appears into ${\cal N}$. 

The minimization condition Eq.(~\ref{mincond2}) determines vacuum expectation
value of the moduli fields
as a function of the electric and the magnetic charges. The condition can
be inverted as follows.
One first recalls the symplectic orthogonality relation Eq.(~\ref{ortho})
of the symplectic covariant vector $U_A$:
\be
\langle U_A | V \rangle = 0 = \langle U_A | {\overline V} \rangle
.
\ee
Then the minimization condition Eq.(~\ref{mincond2}) is solved by a 
{\sl linear map} between the symplectic section $V$ and the electric
and the magnetic charge ${\cal Q}$:
\be
{\cal Q} = c_1 V + c_2 {\overline V}
\ee
where $c_1, c_2$ are complex-valued parameters to be determined. Consistency
condition imposes the symplectic inner product of this solution with $V, 
{\overline V}$ respectively:
\bee
{\bf Z} &\equiv& \langle V | {\cal Q} \rangle \hskip0.5cm \rightarrow 
\hskip0.5cm c_2 = - i {\bf Z}, \nonumber \\
{\overline {\bf Z}} &\equiv& \langle {\overline V} | {\cal Q} \rangle
\hskip0.5cm \rightarrow \hskip0.5cm c_1 = + i {\overline {\bf Z}}.
\eee
Hence, we finally get the K\"ahler-BPS condition:
\bee
{\cal Q} &=& i ({\overline {\bf Z}} V - {\bf Z} {\overline V}), \nonumber \\
P^\Lambda &=& 2 {\rm Im} ({\overline {\bf Z}} L^\Lambda), \nonumber \\
Q_\Lambda &=& 2 {\rm Im} ({\overline {\bf Z}} M_\Lambda).
\label{linrel}
\eee 
Note that the right hand side of Eq.(\ref{linrel}) is invariant under 
the K\"ahler transformation. This is a necessary
condition since the electric and the magnetic charges carry no K\"ahler
weights.

In order to obtain an explicit form of the constant scalar fields, one
needs to solve the K\"ahler BPS condition Eq.(\ref{linrel}).
Using the K\"ahler-BPS saturated quadratic sum rules of the central charges 
Eq.(~\ref{extcentralcharge}), one finds the solution as:
\be
V = - { 1 \over 2 {\overline {\bf Z}}} \Big[ 
\left( \begin{array}{cc} 0&+{\bf I} \hskip-0.16cm {\bf I}\\
- {\bf I} \hskip-0.16cm {\bf I} &0\\
\end{array} \right) \cdot {\bf M}({\cal F}) + i \left( \begin{array}
{cc} + {\bf I} \hskip-0.16cm {\bf I} &0 \\ 0 &+ {\bf I} \hskip-0.16cm {\bf I} 
\\ \end{array} \right) \Big] \cdot {\cal Q}.
\label{minsol}
\ee
On the right-hand side, the first term denotes a particular solution that
satisfies $\langle V | {\cal Q} \rangle = {\bf Z}$, while the second term
is a homogeneous solution that satisfies the K\"ahler-BPS condition
Eq.(~\ref{linrel}).
It is straightforward to check that the solution Eq.(\ref{minsol}) satisfies 
the symplectic constraint $\langle V | {\overline V} \rangle = i$.
Expanding Eq.(\ref{minsol}) in $2n_V$ components, 
\bee
- 2 {\overline {\bf Z}} L^\Lambda \!\!\! &=& \!\!\! \Big[i P
- ({\rm Im} {\cal N})^{-1} ({\rm Re} {\cal N}) \, P
+ ({\rm Im} {\cal N})^{-1} \, Q \Big]^\Lambda
\nonumber \\
-2 {\overline {\bf Z}} M_\Lambda \! \!\! &=& \!\!\!
\Big[i Q
- ( ({\rm Im}) {\cal N}) + ({\rm Re} {\cal N}) ({\rm Im} {\cal N})^{-1}
({\rm Re} {\cal N}) ) \, P
+ ({\rm Re} {\cal N}) ({\rm Im } {\cal N})^{-1} 
\, Q \Big]_\Lambda
\label{minsolcomp}
\eee
one finds an agreement with earlier result by Ferrara and 
Kallosh\cite{ferrarakallosh}.

A comment is in order about the relation  
between the BPS black hole free energy and the topological free energy 
\cite{topfreeenergy} which appears 
naturally in threshold corrections to string effective supergravity. 
In terms of holomorphic sections $\Omega$, the central charge is given:
\be
{\bf Z}({\cal Q}) = e^{K/2} \langle {\cal Q} | \Omega \rangle 
\equiv e^{K/2} {\cal M} ({\cal Q}); \hskip1cm 
        \Omega = (X^\Lambda, F_\Lambda).
\ee
Here, ${\cal M}$ is the so-called holomorphic mass.
We have emphasized that the expression is for a fixed charge vector 
${\cal Q}$ by denoting the dependence on it explicitly.
The topological free-energy is given by
\be
e^{-F_{\rm Top.}} = {\rm Det}_{\{{\cal Q}\}} e^K {\cal M} ({\cal Q}) 
\cdot {\cal M}^\dagger ({\cal Q}),
\label{topfree}
\ee
where the determinant is over the fermionic mass matrix.
Thus the topological free energy sums up contributions of all {\sl virtual} 
BPS states, hence, is an infinite sum over the logarithm of the
 free-energy associated with a single BPS black hole background.
In general, however, minima of the topological free energy is distinct 
from that for the free energy of a single BPS black.
Minima of the BPS black hole free energy is given by the K\"ahler-BPS 
condition
\be
D_A {\bf Z} = 0 \hskip0.5cm \rightarrow \hskip0,5cm 
e^{K/2} \langle {\cal Q} | (\partial_A + K_A) \Omega  \rangle = 0.
\ee
viz. condition for a K\"ahler covariantly constant holomorphic mass:
\bee
{\cal D}_A {\cal M} ({\cal Q})  &\equiv& 
(\partial_A + K_A) {\cal M} ({\cal Q}) = 0,
\nonumber \\
\rightarrow \hskip0.75cm \partial_A \log {\cal M} ({\cal Q}) &=& 
- \partial_A K ({\cal Q}).
\label{bhfreecond}
\eee
On the other hand, the minima of topological free energy is 
determined by:
\be
\partial_A F_{\rm Top.} = \sum_{\{{\cal Q} \}} [
\partial_A \log {\cal M}({\cal Q}) + \partial_A K] = 0.
\label{topfreecond}
\ee
In Eq.(\ref{topfreecond}), while the summand equals to the 
condition Eq.(\ref{bhfreecond}),
it does not necessarily require for each term in the summand  to vanish.
It is evident that minima of the topological free energy
is generically different from that determined by the K\"ahler-BPS condition.
\subsubsection{\rm K\"AHLER-BPS CONDITION IN THE SHIFTED BASIS}
It is possible to simplify the solution given in 
Eq.(~\ref{minsol},~\ref{minsolcomp}) 
further by making a symplectic transformation that 
amounts to the Witten effect~\cite{witteneffect} and associated shift 
of the electric charge. One first recalls that the quadratic sum rule of the 
central charge becomes manifestly a positive-definite quadratic form 
once the K\"ahler-BPS condition is satisfied:
\bee
|{\bf Z}|^2
&=& - {1 \over 2} {\cal Q}^T \cdot {\bf M} ({\cal F}) \cdot {\cal Q},
\nonumber \\
{\bf M}({\cal F}) &=& {\cal R}^T ({\rm Re} {\cal F})
\cdot {\cal D} ({\rm Im} {\cal F}) \cdot {\cal R} ({\rm Re} {\cal F}).
\eee
Define the following new symplectic sections and symplectic 
charges\footnote{As mentioned at the end of Section2.2, one 
can define another symplectic transformed vectors and charges 
in which the holomorphic matrix ${\cal F}$ in the following equations 
is replaced by the coupling matrix ${\cal N}$.}:
\bee
{\bf V} & \equiv & \left( \begin{array}{c}
{\bf L}^\Lambda \\ {\bf M}_\Lambda  \end{array} \right) 
= {\cal R} ({\rm Re} {\cal F}) \cdot V
= \left( \begin{array}{cc} {\bf I} \hskip-0.16cm {\bf I} & 0 \\ 
-{\rm Re} {\cal F} & {\bf I} \hskip-0.16cm {\bf I} \\ \end{array} 
\right) \cdot 
\left( \begin{array}{cc} L^\Lambda \\ M_\Lambda \\ \end{array} 
\right), \label{shiftsection} \\
{\bf Q} & \equiv & \left( \begin{array}{c}
{\bf p}^\Lambda \\ ({\rm Im} {\cal F})_{\Lambda \Sigma} \cdot {\bf q}^\Sigma 
\\ \end{array} \right)
= {\cal R} ({\rm Re} {\cal F}) \cdot {\cal Q} 
= \left( \begin{array}{cc} {\bf I} \hskip-0.16cm {\bf I} & 0 \\ 
- {\rm Re} {\cal F} & {\bf I} \hskip-0.16cm {\bf I} \\ \end{array} 
\right) \left( \begin{array}{c} P^\Lambda \\ Q_\Lambda
\end{array} \right).
\label{shiftcharge} 
\eee
A few comments are in order for the electric and the magnetic charges
in the new shifted basis. First, while the fundamental electric and
magnetic charges of the symplectic charge vector ${\cal Q} = (P^\Lambda,
Q_\Lambda)$ are integer-valued, hence, independent of the special coordinates
$Z^A$, the shifted electric charge $({\rm Im} {\cal F}) \cdot {\bf q}$
after the symplectic transformation depends on the special coordinates,
hence, takes an arbitrary value. This is the manifestation of induced charge
effect both due to the screening from ${\rm Im} {\cal N}$ and due to the 
Witten effect\footnote{ This difference is also reflected on the 
Dirac-Schwinger-Zwanziger quantization condition of ${\bf p}, {\bf q}$ 
charges:
\be
{\rm Im} {\cal F}_{\Lambda \Sigma} ({\bf p}^\Lambda {\bf q}^{'\Sigma}
- {\bf p}^{'\Lambda} {\bf q}^\Sigma) = (2 \pi \hbar) n .
\ee
}.          The shifted magnetic charge ${\bf p}^\Lambda$, however, 
remains unchanged ${\bf p}^\Lambda  = P^\Lambda$.
Second, the shifted charges ${\bf Q} = ({\bf p}^\Lambda, {\bf q}^\Lambda)$
in the new basis are precisely the ones that appear in the gauge field 
strength of the BPS black hole solution Eq(\ref{bhansatz}), viz. charges
that are measured outside the horizon.

One also notes that the constitutive relations for the new symplectic
sections is given by\footnote{
In deriving this relations we have used Eq.(2.30) and the fact that 
${\cal N} \cdot L = {\cal F} \cdot L$, which can be checked directly 
from Eq.(2.33).}
\bee
{\bf M}_\Lambda &=& [ - ({\rm Re} {\cal F}) \cdot L + M]_\Lambda 
\nonumber \\
&=& [ - ({\rm Re} {\cal F }) \cdot L + {\cal F} \cdot L]_\Lambda \nonumber \\
&=& i \, ({\rm Im} {\cal F})_{\Lambda \Sigma} {\bf L}^\Sigma.
\eee

For K\"ahler-BPS states, the quadratic form of the central charge reads 
in the new shifted basis as:
\bee
|{\bf Z}|^2 &=& - {1 \over 2} {\bf Q}^T \cdot {\cal D} 
({\rm Im}  {\cal F}) \cdot {\bf Q} 
\nonumber \\
&=& -{1 \over 2} \Big[ {\bf p}
\cdot {\rm Im} {\cal F} \cdot {\bf p} + {\bf q} \cdot {\rm Im} {\cal F} 
\cdot {\bf q} \Big]
\label{minZ}
\eee

Since the K\"ahler-BPS solution Eq.(\ref{minsol}) is symplectically covariant, one may simply replace the original sections and charges into the new 
shifted ones so that the form of equation is unchanged\footnote{One can 
derive this new, shifted K\"ahler-BPS solution directly from Eq.
(\ref{minsol}) using Eq.(\ref{rsymp}).} : 
\be
{\bf V} = - {1 \over 2 {\overline {\bf Z}}} \Big[
\left( \begin{array}{cc} 0&+{\bf I} \hskip-0.16cm {\bf I} \\ 
- {\bf I} \hskip-0.16cm {\bf I} & 0 \\ \end{array} \right) \cdot
{\cal D}({\rm Im} {\cal F}) + i \left( \begin{array}{cc} 
+{\bf I} \hskip-0.16cm {\bf I} & 0 \\ 0 & +{\bf I} \hskip-0.16cm {\bf I} \\
\end{array} \right) \Big] \cdot {\bf Q}.
\label{newcond}
\ee
It is easy to check that the $(2n_V + 2) \times (2n_V + 2)$ matrix inside 
the bracket on the right hand side of Eq.(~\ref{newcond}) has a rank 
$(n_V + 1)$ only. This is as it should be since the $L^\Lambda$ and the 
$M_\Lambda$ are 
related each other for a given $(2n_V + 2)$ electric and magnetic charges. 
Therefore, it is enough to solve only half components of Eq.(\ref{newcond}).
Typically since the $L^\Lambda$ sections are not modified by the shift
symplectic transformation, it is more convenient to solve them.
In terms of the shifted holomorphic sections, this is easily seen: 
\bee
{\bf L}^\Lambda &=& L^\Lambda = {\rm M}_{\rm Pl} e^{K/2} X^\Lambda,
\nonumber \\
{\bf M}_\Lambda &=& i ({\rm Im} {\cal F})_{\Lambda \Sigma} {\bf L}^\Sigma = i
({\rm Im} {\cal F})_{\Lambda \Sigma} L^\Sigma.
\eee
To keep the Einstein-Hilbert term in the $N=2$ supergravity 
Lagrangian, it is necessary to choose the $X^0 = 1$ gauge. 
This gauge choice then determines the K\"ahler-BPS central charge in terms
of the K\"ahler potential once the electric and magnetic charges are 
specified:
\be
- 2 {\rm M}_{\rm Pl} {\overline {\bf Z}} e^{K/2} = [i {\bf p} + {\bf q}]^0.
\label{xogauge}
\ee
Therefore, the K\"ahler-BPS black hole mass and the macroscopic entropy is 
given:
\be
{\bf M}_{\rm BPS}^2 = {\rm M}_{\rm Pl}^2 \Big( 
{{\rm S}_{\rm BH} \over \pi} \Big) = 
{\rm M}_{\rm Pl}^2 | {\bf Z}|^2 = {1 \over 4} 
e^{-K} [({\bf p}^0)^2 + ({\bf q}^0)^2].
\label{bpscharge}
\ee

Using the gauge fixing relation Eq.(\ref{xogauge}), 
the $n_V$ special coordinates that saturate the K\"ahler-BPS bound  
can be expressed as a homogeneous rational function of the charges:
\be
Z^A = \Big( {{\bf L}^A \over {\bf L}^0} \Big) = \Big({X^A \over X^0} \Big) = 
{[i {\bf p} + {\bf q} ]^A
\over [i {\bf p} +  {\bf q}]^0 }.
\ee
On the other hand,
\be
\Big( { {\bf M}_\Lambda  \over {\bf L}^0} \Big) 
= - ({\rm Re} {\cal F})_{\Lambda \Sigma} Z^\Sigma + 
\Big({F_\Lambda \over X^0} \Big)
= i ({\rm Im} {\cal F})_{\Lambda \Sigma} { [ i {\bf p} + {\bf q}]^\Sigma
\over  [i {\bf p} +  {\bf q}]^0 }.
\ee
The first equality clearly shows a generalization of the Witten effect.
By the exactly same argument given in section 2.1.2 for the rigid $N=2$ 
case, one may first shift the generalized vacuum angle 
${\rm Re} {\cal F}$ to zero by the symplectic transformation and minimize 
the ${\rm Im} {\cal F}$. Subsequently the ${\rm Re} {\cal F}$ may be
re-introduced by un-doing the symplectic transformation.
As shown in the rigid $N=2$ case, this procedure is essentially equivalent
to maintaining the symplectic section $M_\Lambda$ or, equivalently, 
the holomorphic section $F_\Lambda$ unshifted in the new basis defined by 
the symplectic transformation Eq.(~\ref{shiftsection}). This method will
be adopted when we solve the K\"ahler-BPS conditions and determine the
special coordinates explicitly for the heterotic string theory.

\subsection{Example: N=4 Supergravity}
To demonstrate the utility of the shifted symplectic basis, consider the 
$N=4$ supergravity described as the $N=2$ supergravity coupled to 
two vector fields \cite{kalloshstudent1}.  The prepotential of vector fields is
given by $F = -i X^0 X^1$. 
Calculating the holomorphic matrix ${\cal F}_{\Lambda 
\Sigma } \equiv \partial_\Lambda \partial_\Sigma F(X)$:
\be
{\cal F} = \left( \begin{array}{cc} 0 & - i \\ - i & 0 \\ \end{array}
\right) ; \hskip1cm {\cal F}^{-1} = \left( \begin{array} {cc}
0 & +i \\ +i & 0 \\ \end{array} \right).
\ee
One finds immediately the special coordinate for K\"ahler-BPS states: 
\be
Z \equiv \Big({X^1 \over X^0} \Big)  = {[ {\bf p}^1 + i {\bf q}^1 ] 
\over [{\bf p}^0 + i {\bf q}^0]} = {[i P^1 - Q_0] \over
[i P^0 - Q_1]}.
\ee
The K\"ahler-BPS black hole mass and the entropy is given by
\bee
{\bf M}_{\rm BH}^2 = {\rm M}_{\rm Pl}^2 \big( {{\rm S}_{\rm BH} \over \pi} 
\big) &=& 
{\rm M}_{\rm Pl}^2 |{\bf Z}|^2 = - {1 \over 2} {\bf Q}^T
 \cdot 
\left( \begin{array}{cc} {\rm Im} {\cal F} & 0 \\ 0 & ({\rm Im} {\cal F})^{-1} 
\\ \end{array} \right) \cdot {\bf Q} \nonumber \\
&=& [{\bf p}^0 {\bf p}^1 + {\bf q}^0 {\bf q}^1]^2 
\nonumber \\
&=& [P^0 P^1 + Q_0 Q_1]^2.
\eee 
Applying the following symplectic transformation, 
\bee
{\hat \Omega}: \,\,\, (X^0, X^1, F_0, F_1) \hskip0.5cm  &\rightarrow& 
\hskip0.5cm ({\hat X}^0, {\hat X}^1, {\hat F}_0, {\hat F}_1 ) 
\nonumber \\
  \,\,\, (P^0, P^1, Q_0, Q_1) \hskip0,5cm &\rightarrow& \hskip0.5cm 
({\hat P}^0, {\hat P}^1, {\hat Q}_0, {\hat Q}_1 ),
\eee
where 
\be
{\hat \Omega} = \left( \begin{array}{cccc}
+1 & 0 & 0 & 0 \\
0 & 0 & 0 & -1 \\
0 & 0 & +1 & 0 \\
0 & +1 & 0 & 0 \\
\end{array} \right).
\ee
one obtains the $SU(4)$ formulation of the $N=4$ supergravity.
Since the matrix ${\cal D}({\rm Im} {\cal F})$ transforms as
\be
{\hat \Omega}: \,\,\,
{\cal D} \hskip0.5cm \rightarrow \hskip0.5cm 
{\hat \Omega} \cdot {\cal D} \cdot {\hat \Omega}^T = 
\left( \begin{array}{cccc} 
0 & 0 & 0 & +1 \\
0 & 0 & -1 & 0  \\
0 & -1 & 0 & 0 \\
+ 1 & 0 & 0 & 0 \\
\end{array} \right),
\ee
one finds the K\"ahler-BPS black hole mass and entropy in the $SU(4)$
basis as: 
\be
{\bf M}_{\rm BH}^2 = {\rm M}_{\rm Pl}^2 \big( {{\rm S}_{BH} \over \pi} \big)
= {\rm M}_{\rm Pl}^2 |{\bf Z}|^2 = 
[{\hat {\bf p}}^0 {\hat {\bf q}}^1 - {\hat {\bf p}}^1 {\hat {\bf q}}^0]^2 =
[{\hat P}^1 {\hat Q}_0 - {\hat Q}_0 {\hat P}^1 ]^2.
\ee
\section{HETEROTIC EXTREME BLACK HOLES: CLASSICAL ASPECTS} 
\setcounter{equation}{0}
\subsection{Rank-3 Heterotic Compactification: The STU Model}
The simplest yet nontrivial $D=4, N=2$ heterotic string vacua is obtained 
from from further $T_2$ compactification of the $D=6, N=1$ heterotic string 
compactified on $K_3$ with instanton number $(12,12)$~\footnote{
For extensive study of heterotic $K_3$ compactification, see
~\cite{ibanez}.}. At generic 
point in the moduli space, gauge symmetry is completely Higgsed
and one is left with supergravity multiplet and one tensor multiplet.
Further compactification on $T_2$ then
gives rise to rank-3 gauge groups\footnote{This is the rank of the gauge 
group not including the gravi-photon.} in four dimensions, of which one 
is associated with the heterotic dilaton vector multiplet and the other two 
with the $T, U$ moduli vector multiplets.
It is now well-established~\cite{kachruvafa}~\cite{furthertest}
that this, so-called STU-model
is dual to the $D=4, N=2$ type IIA string compactification
on a Calabi-Yau three-fold defined by a weighted-projective space 
${\bf P}_{1,1,2,8,12}(24)$. 
It is also known~\cite{seiberggroup} to be dual 
to $T_2$ compactification of $D=6$ type-I orientifold
model on $K_3$ orbifold $T_4 / {\bf Z}_2$ with one tensor multiplet
and completely Higgs gauge group.
On the type IIA side, the classical prepotential in the large K\"ahler 
volume limit is given by
\bee
F_II &\equiv& d_{ABC} X^A X^B X^C / X^0
\nonumber \\
&=& (X^0)^2 STU
\eee
where $d_{ABC} \equiv \int J_A \wedge J_B \wedge J_C$ ($J_A$'s are generators
of K\"ahler cone) denotes the classical intersection numbers and the 
heterotic special coordinates are used in the last expression.
The prepotential displays an explicit STU triality associated with 
the special K\"ahler manifold ${\cal M}_{SK} = [SU(1,1)/U(1)]_S 
\times [SO(2,2)/ SO(2) \times SO(2)]_{T,U}$.
By type IIA-heterotic string-string duality map, one obtains the
heterotic STU model, for which the heterotic dilaton $S$ is picked up 
as a special polarization in the K\"ahler moduli space. 
On the heterotic side, it is known that~\cite{ferraraetal}
 one needs to make a symplectic
transformation corresponding to a strong-weak coupling exchange in
order to yield a uniform weak coupling behavior as 
${\rm Im} S \rightarrow \infty$. 
This symplectic transformation is defined as: 
\be
X^1 \rightarrow F^{(0)}_1; \hskip1cm F^{(0)}_1 \rightarrow - X^1.
\label{strongweak}
\ee
In this case, one finds that the holomorphic section is given by 
\be
\Omega = \left( \begin{array}{c} X^\Lambda \\ F_\Lambda \end{array} \right)
= \left( \begin{array}{c} X^\Lambda \\ S \eta_{\Lambda \Sigma} X^\Sigma
\end{array} \right),
\ee
where
\be
\eta_{\Lambda \Sigma} = \left( \begin{array}{cccc} 
\, 0\,  & \, 1 \,  & \, 0 \, & \, 0 \, \\
\, 1 \, & \, 0 \, & \, 0 \, & \, 0 \, \\
\, 0 \, & \, 0 \, & \, 0 \, & \, 1 \, \\
\, 0 \, & \, 0 \, & \, 1 \, & \, 0 \, \\
\end{array} \right); \hskip0.75cm \eta^2 = {\bf I}.
\ee
In special coordinates,
\bee
X^\Lambda &=& (1, - TU, T, U); 
\nonumber \\
F^{(0)}_\Lambda &=& (- STU, S, SU, ST).
\eee 
One now finds that the new sections are not mutually independent but are 
constrained as:
\be
\langle X, X \rangle = 0 \hskip1cm {\rm where} \hskip1cm
\langle A, B \rangle
\equiv A^\Lambda \eta_{\Lambda \Sigma} B^\Sigma.
\label{constraint}
\ee
It is important to realize that this constraint has to be obeyed not only 
at classical level but also at quantum level. 
This is because the constraint has stemmed from the geometrical aspect 
that one of the original sections, $X^1$, that are needed for parametrizing
freely the special K\"ahler manifold ${\cal M}_{SK}$ is lost\footnote{
Because of this $SO(2,2)$ is realized nonlinearly in the new basis.} by the
symplectic transformation Eq.(\ref{strongweak}) and from the fact that the
heterotic dilaton $S$, which defines the string-loop counting parameter
resides only to the $F_\Lambda$ sections.  
One also finds that $\langle F^{(0)}, F^{(0)} \rangle = 0$. However, 
this relations is not expected to be valid beyond the classical approximation, 
since  the sections $F_\Lambda$ depend on the heterotic dilaton $S$ already
at the classical level, hence, subject to quantum corrections. 

The K\"ahler potential of the heterotic STU model is easily found:
\bee
K &=& - \log [ i \langle \Omega | {\overline \Omega} \rangle ]
\nonumber \\
&=& - \log[ 2 {\rm Im} S] - \log[ \langle X, {\overline X} \rangle]
\nonumber \\
&=& - \log[ 2 \, {\rm Im} S ] - \log[ 4 \, {\rm Im} T \, {\rm Im} U].
\label{kahlerpot}
\eee
One finds the following K\"ahler metric at classical level:
\bee
K_{S {\overline S}} &=& {1 \over 4 ( {\rm Im} S)^2}, \nonumber \\
K_{\Lambda \Sigma} &=& - {1 \over \langle X , {\overline X} \rangle }
\Big[ \eta_{\Lambda \Sigma} - {{\overline X}_\Lambda X_\Sigma 
+ {\overline X}_\Sigma X_\Lambda \over2 \langle X , {\overline X} \rangle
} \Big];
\nonumber \\
\rightarrow \hskip0.5cm 
K_{T {\overline T}} &=& { 1 \over 4 ({\rm Im} T)^2}, \hskip0.75cm 
K_{U {\overline U}} = {1 \over 4 ({\rm Im} U)^2}.
\eee
One also finds the holomorphic matrix as:
\bee
{\cal F}^{(0)}_{\Lambda \Sigma} \equiv \partial_\Lambda F^{(0)}_\Sigma 
&=& S \eta_{\Lambda \Sigma} ; \hskip1.75cm
\Big[ {1 \over {\cal F}^{(0)} } \Big]^{\Lambda \Sigma} 
= {1 \over S} \eta^{\Lambda \Sigma}
, \nonumber \\
({\rm Im} {\cal F}^{(0)} )_{\Lambda \Sigma} 
&=& ({\rm Im} S) \eta_{\Lambda \Sigma}; 
\hskip0.75cm  
\Big[ {1 \over {\rm Im} {\cal F}^{(0)} }]^{\Lambda \Sigma}  
= ({\rm Im} S)^{-1} \eta^{\Lambda \Sigma}.
\label{hetF}
\eea
From this one finds the metric ${\cal D}({\rm Im} {\cal F})$ that defines 
the quadratic form of the BPS mass spectra Eq.(\ref{minZ}): 
\be
{\cal D}({\rm Im} {\cal F}) = 
\left( \begin{array}{cc}
({\rm Im} S) \,  \eta & 0 \\
0 & ({\rm Im} S)^{-1} \, \eta \\
\end{array} \right).
\ee
One notes that the (classical) heterotic 
dilaton ${\rm Im} S$ sets a universal coupling parameter to all
gauge fields, hence, the BPS mass spectra and the central charge 
take precisely the same form as those of the rigid $N=2$ supersymmetric 
gauge theories studied in section 2.1.3.
One thus expects that the K\"ahler-BPS conditions of the heterotic
string theory follow a similar pattern as those in the rigid $N=2$ theories,
Eqs.(\ref{extangle}, ~\ref{extcoupl1}). In the next section, we will find that 
this turns out the case.

The $T, U$ special coordinates are the moduli fields that parametrize
inequivalent ground state of spontaneously broken rank-2 gauge group. 
It is known that the $U(1) \times U(1)$ gauge group 
associated with $T, U$ fields
are enhanced to $SU(2) \times U(1)$ at $T=U \ne 1$ complex curve, to
$SU(2) \times SU(2)$ at $T=U=1$ point and to $SU(3)$ at $T = 1/U = 
\exp(i \pi / 6)$ point.
Thus, the values of $T, U$ fields away from these enhanced gauge symmetry
points and curve can be interpreted as the string counterpart of the 
vacuum expectation value of Higgs fields in the rigid $N=2$ gauge theory.
In fact, it is known that~\cite{narain}, 
much as the Higgs expectation value $v$ in the latter theory
sets the mass scale of the elementary excitations in the theory such as 
the heavy charged gauge boson mass ${\tt M}_W = g v$, 
the vacuum expecation values of $T, U$ fields sets the mass scale of the
elementary string excitations such as the Kaluza-Klein and the winding string
states that arise upon compactification on $T_2$.
Likewise, the mass scale of non-perturbative soliton excitations such as 
magnetic monopoles and dyons in the rigid $N=2$ theory and black holes in the
local $N=2$ theory is also determined by the same vacuum expectation values of 
the Higgs field and of the $T, U$ fields respectively.
\subsection{Dynamical Relaxation of Heterotic Black Hole Mass and Entropy}
One now solve the classical K\"ahler-BPS condition:
\be
{\bf V} = - { 1 \over 2 {\overline {\bf Z}}} 
\left( \begin{array}{cc}
i {\bf I} \hskip-0.16cm {\bf I} & ({\rm Im} S)^{-1} \cdot \eta \\
- {\rm Im} S \cdot \eta & i {\bf I} \hskip-0.16cm {\bf I} \\
\end{array} \right) \cdot {\bf Q}.
\ee
Expanding the components in terms of holomorphic sections, one gets
\bee
- 2 {\overline {\bf Z}} {\rm M}_{\rm Pl} e^{K/2} \, X^\Lambda
&=& [ i {\bf p} + {\bf q} ]^\Lambda, \nonumber \\
 -2 {\overline {\bf Z}} {\rm M}_{\rm Pl} e^{K/2} \, S \,
 (\eta \cdot X)_\Lambda
&=& i {\rm Im} S \, [ \eta \cdot (i {\bf p} + {\bf q})]_\Lambda.
\label{exteqns}
\eee
As explained in section 2.4.2 on a general ground, one finds that the 
second set of equations in Eq.(\ref{exteqns}) is identical to those of
the first set. One now solves the first set of equations to determine the 
K\"ahler-BPS configurations of $S, T, U$ fields as well as the 
BPS black hole mass and entropy.

Since the $S$-field sets the (classical) heterotic string coupling
parameters, one first determine K\"ahler-BPS configuration of this
field.
It is determined by the requirement that the K\"ahler-BPS configurations
of the scalars respect the geometric constraint Eq.(\ref{constraint}):
\be
\langle X, X \rangle = 0 \hskip0.5cm \rightarrow \hskip0.5cm
\langle (i {\bf p} + {\bf q}), (i {\bf p} + {\bf q}) \rangle= 0.
\ee 
Decomposing this constraint into real and imaginary parts,
one obtains
conditions
\bee
\langle {\bf p}, {\bf p} \rangle &=&  \langle {\bf q}, {\bf q}
\rangle, \label{Re} \\
({\bf p} \cdot {\bf q}) &=& 0.
\label{Im}
\eee
Recalling
the definition of shifted charged vector ${\bf Q} = ({\bf p}, {\bf q})$
given in Eq.(\ref{shiftcharge}) and the structure of the classical 
holomorphic matrix Eq.(\ref{hetF}), one notes that these two constraints
are precisely of the same conditions Eqs.(\ref{extangle},~\ref{extcoupl1})
needed for the minimized the BPS mass spectra in the rigid $N=2$ gauge 
theories.

Following exactly the same steps as in the rigid case, 
one first solves the Eq.(\ref{Im}) and determine ${\rm Re} S$-field 
configuration in terms of the microscopic charges:
\bee
({\bf p} \cdot {\bf q}) &=& ({1 \over {\rm Im} S}) \, [ - ({\rm Re} S)
\langle P, P \rangle + (P \cdot Q)  ] = 0;
                           \nonumber \\
\rightarrow \hskip0.75cm {\rm Re} S &=& {(P \cdot Q) \over \langle
P, P \rangle }.
\eee
Using this result, one also solves the Eq.(\ref{Re}) and
obtain the ${\rm Im} S$-field configuration as:
\bee
\langle P, P \rangle = \langle {\bf p}, {\bf p} \rangle
&=& \langle {\bf q} , {\bf q} \rangle
= {1 \over ({\rm Im} S )^2} \,
\langle (Q - {(P \cdot Q) \over \langle P, P \rangle} P), 
( Q - { (P \cdot Q) \over \langle P, P \rangle } P) \rangle ;
\nonumber \\
\rightarrow \hskip0.75cm 
{\rm Im} S &=&
\sqrt {{\langle Q, Q \rangle \over \langle P, P \rangle} - {
(P \cdot Q) \over \langle P, P \rangle }}.
\eee

Altogether, we have the K\"ahler-BPS configuration of the $S$-field:
\be
S \equiv [{\theta \over 2 \pi} + i {4 \pi \over g^2}]_{\rm Cl}
= {(P \cdot Q) \over \langle P, P \rangle }
 + i \sqrt { {\langle Q, Q \rangle \over \langle P, P \rangle}
- {(P \cdot Q) \over \langle P, P \rangle} }.
\ee

K\"ahler-BPS configuration of the moduli fields $T, U$ are obtained as
a homogeneous rational functions of charges:
\bee
T &=& \Big( {X^2 \over X^0} \Big) = {[i {\bf p} + {\bf q}]^2
\over [i {\bf p} + {\bf q}]^0},
\label{t} \\
U &=& \Big( {X^3 \over X^0} \Big) = { [i {\bf p} + {\bf q} ]^3 \over
[i {\bf p} + {\bf q}]^0}.
\label{u}
\eee
In addition, one also finds
\be
- TU = \Big( {X^1 \over X^0} \Big)
= { [ i {\bf p} + {\bf q}]^1 \over [i {\bf p} + {\bf q}]^0 }.
\label{tu}
\ee
That this equals to the product of the two expressions of Eqs.(\ref{t},
\ref{u})
is guaranteed by the constraint:$ \langle X, X \rangle = 0$. One can
verify this by direct multiplication of Eqs.(\ref{t},~\ref{u}) and
comparison with Eq.(\ref{tu}) using the charge relations 
Eqs.(\ref{Re},~\ref{Im}).

A consistency check of the above K\"ahler-BPS configuration is 
provided by deriving the BPS black hole mass and entropy explicitly.
Taking an inner product of the
K\"ahler-BPS configuration Eq.(\ref{exteqns}) with a complex-conjugate
of itself and using the Eq.(\ref{Im}), one obtains 
\be
4 {\rm M}_{\rm Pl}^2 e^K \langle X, {\overline X} \rangle |{\bf Z}|^2 
= [ \langle {\bf p}, {\bf p} \rangle + \langle {\bf q}, {\bf q} \rangle ].
\label{bpsexp}
\ee
Using the form of the K\"ahler potential given in Eq.(\ref{kahlerpot})
one finds the K\"ahler-BPS black hole mass and the entropy:
\bee
{\bf M}_{\rm BPS}^2 = {\rm M}_{\rm Pl}^2 \Big({S_{\rm BH} \over \pi} \Big)
= {\rm M}_{\rm Pl}^2 |{\bf Z}|^2 &=& 
{1 \over 2} ({\rm Im} \, S) 
[ \langle {\bf p}, {\bf p} \rangle + \langle {\bf q}, {\bf q} \rangle]
\ge ({\rm Im} S) \, \sqrt { \langle {\bf p}, {\bf p} \rangle
\langle {\bf q}, {\bf q} \rangle }
\nonumber \\
& = & ({\rm Im} S) \, \langle {\bf p} , {\bf p} \rangle
= ({\rm Im} S) \, \langle P, P \rangle
\nonumber \\
&=& \sqrt { \langle P, P \rangle \langle Q, Q \rangle - (P \cdot Q)^2}.
\eee
At the end of the first line, we have indicated the Cauchy-Schwarz inequality
valid generally for the quantity of the preceding expression. 
To obtain the second line, we have used the condition Eq.(\ref{Re}) to the
middle expression of the first line. Noting that this equals the 
value of the last expression in the first line, one concludes that the 
geometric constraints Eqs.(\ref{Re},\ref{Im}) are nothing but
the condition for saturating the Cauchy-Schwarz inequality.
Note that the condition for saturating the H\"older's inequality
Eq.(\ref{Im}) has been used already in obtaining Eq.(\ref{bpsexp}).

\section{HETEROTIC EXTREME BLACK HOLES: QUANTUM ANALYSIS}
\setcounter{equation}{0}
\subsection{Quantum BPS Mass and Dynamical relaxation in Rigid $N=2$ Theory}
\subsubsection{\rm RIGID $N=2$ NON-RENORMALIZATION THEOREMS}
We first recall the known results for the non-renormalization 
theorems in rigid $N=2$ supersymmetric gauge theories~\cite{barbieri}
\footnote{ The results were subsequently extended to coupling to 
supergravity~\cite{barbieri} with consistent regularization 
and renormalization prescriptions~\cite{gaillard}.}
The one-loop radiative correction is most straightforwardly calculated 
in the background field method by evaluating determinants of Gaussian
fluctuations around a fixed background of bosons, fermions and Faddeev-Popov 
ghosts. In terms of $N=1$ supermultiplets, denoting the wave function 
renormalizations of the vector field, 
the scalar fields in the adjoint representation and chiral matter fields in 
the complex-conjugate pair representations as ${\tt Z}_V, {\tt Z}_A, 
{\tt Z}_Q, {\tt Z}_{\overline Q}$, the gauge coupling renormalization
as ${\tt Z}_g$ and the mass of adjoint scalar and complex-conjugate pair
scalars as ${\tt M}_W$ and ${\tt M}_Q$, 
it was found~\cite{barbieri} that they are related each other as: 
\bee
{\tt Z}_g \cdot [ {\tt Z}_V]^{1 \over 2} &=& 1, \nonumber \\
{\tt Z}_g \cdot {\tt Z}_Q^{1\over 2} 
\cdot {\tt Z}_{\overline Q}^{1 \over 2} {\tt Z}_A &=& 1, \nonumber \\
{\tt Z}_Q^{1 \over 2} 
{\tt Z}_{\overline Q}^{1\over 2} [{\tt M}_Q]_{\rm Bare}
&=& [{\tt M}_Q ]_{\rm Ren}.
\eee

For $N=2$ supersymmetric gauge theories, the requisite $N=2$ vector and
hyper multiplet structures impose two conditions 
${\tt Z}_V = {\tt Z}_A$ and ${\tt Z}_Q = {\tt Z}_{\overline Q}$ 
respectively.
Therefore, one finds the following two $N=2$ nonrenormalization theorems:
\bee
 {\tt Z}_g \cdot {\tt Z}_V^{1/2} &=& 1 \hskip0.75cm \rightarrow 
\hskip0.75cm [{\tt M}_W]_{\rm Ren} = [{\tt M}_W ]_{\rm Bare}, 
\label{wmass} \\
 {\tt Z}_Q \cdot {\tt Z}_{\overline Q} &=& 1 
\hskip0.75cm \rightarrow \hskip0.75cm 
[{\tt M}_Q]_{\rm Ren} = [{\tt M}_Q ]_{\rm Bare},
\label{hypermass}
\eee
protecting the spectra of the massive vector and hyper multiplets from 
renormalization. While the non-renormalization theorem was derived
for the logarithmic corrections, it should hold also for finite 
renormalizations including threshold corrections due to heavy particles.

Similar non-renormalization theorem applies to the $N=2$ BPS monopoles
and dyons. One might naively expect that the BPS masses are not renormalized
at all since the BPS spectra is determined by the central charge that has 
a topological origin, see Eq.(\ref{centralch}). A heuristic argument
was that the Gaussian fluctuation spectra of bosons and fermions around 
any supersymmetric configuration are equal each other by supersymmetry, 
hence, a sum over the fluctuation 
energy cancels out between the bosonic and the fermionic contributions
~\cite{wittenolive, divecchia}.
However, for $N=2$ supersymmetric theories, this turned out not to be 
the case~\cite{kaul}. A subtle but important point was that the Gaussian fluctuations 
around the BPS soliton configuration contain not only discrete, bound 
states but also a continuum scattering states. 
Thus, schematically, quantum correction to the BPS soliton mass is
given by:
\be
[\Delta {\bf M}]_{\rm 1-loop} 
= \sum \hskip-0.45cm \int \hbar \Omega_{\rm boson}
 - {1 \over 2} \sum \hskip-0.45cm \int \hbar \Omega_{\rm fermion}
- {1 \over 2} \sum \hskip-0.45cm \int \hbar \Omega_{\rm ghost}
.
\ee
For the continuum contributions, while bosonic and fermionic spectra
are always paired as dictated by the supersymmetry, the density of 
states turns out not equal for the bosonic and the fermionic parts
~\footnote{ The BPS monopole mass of $N=4$ supersymmetric gauge theory 
is not renormalized since the difference of density of states between
bosons and fermions turns out to vanish identically for the 
$N=4$ multiplet spectra.}
Explicit one-loop calculations for $N=2$ BPS monopole have shown that 
the BPS mass receives a logarithmic radiative correction
\bea
[{\bf M}_{\rm BPS} ]_{\rm 1-loop} 
= [{\bf M}_{\rm BPS}]_{\rm bare} \Big[
1 - \big( {g^2 \over 4 \pi } \big) {\hbar \over \pi} 
\, \log ({2 \Lambda \over M_W})^2 \Big]
\label{monopolemasscorr}
\eea
where $[{\bf M}_{\rm BPS}]_{\rm bare} = 4 \pi v / g$ and
${\tt M}_W = g v$ denotes the heavy charged gauge boson mass.
At same order, one also finds one-loop radiative correction to the 
gauge coupling constant
 \bee
\Big({4 \pi \over g^2} \Big)_{\rm 1-loop} = 
\Big( {4 \pi \over g^2} \Big)_{\rm bare} \Big[ 1 - \big({g^2 \over 4 \pi}
\big) \, { \hbar \over \pi} \, \log ( {2 \Lambda \over M_W})^2 \Big].
\label{couplingcorr}
\eea
Comparing Eq.(\ref{monopolemasscorr}) with Eq.(\ref{couplingcorr})
one finds that radiative correction to the BPS monopole mass
is entirely due the radiative correction to the gauge coupling constant
that the BPS monopole mass depends on.
One now use the $N=2$ non-renormalization theorem Eq.(\ref{wmass})
and re-express the quantum BPS monopole mass as: 
\bee
[{\bf M}_{\rm BPS}]_{\rm ren} &=& \Big({ 4 \pi v \over g} \Big)_{\rm ren}
\nonumber \\
&=&
 {[ g \, v ]_{\rm ren} \over \alpha_{\rm g, ren} }
= { [ g \, v ]_{\rm bare} \over \alpha_{\rm g, ren}}
\nonumber \\
&=& {{\tt M}_W \over \alpha_{\rm g, ren}}.
\eee
The above analysis indicates that, while the $N=2$ BPS spectra 
receives nontrivial radiative corrections, 
the BPS mass spectra takes exactly the same form once
all the quantities are expressed in terms of renormalized physical
quantities. In particular, if there are no massless charged states
in the elementary spectra, then the renormalization effect will be 
dominated by threshold corrections.
It is in this case that the notion of BPS states as those of 
balancing long-distance static force between them retains a well-defined 
meaning at quantum level. All the short-distance details are summarized
into the renormalization effects to physical parameters and the
renormalization effects are exponentially suppressed at a distance
larger than the typical Compton wavelength of the heavy charged states.

It is then a natural question to what extent the above results for
the quantum BPS spectra for rigid supersymmetric case extends to hold
once they are embedded into supergravity theory. 

We now turn to this question by embedding the gauge theories into
heterotic string theory.

\subsection{Perturbative Quantum Effect: Heterotic $STU$ Model}
Consider perturbative string-loop corrections to the K\"ahler-BPS black hole 
configuration, mass and entropy formula of the heterotic STU model studied 
in section 3.2. In this section, we study the perturbative string-loop effects
to the classical K\"ahler-BPS black hole by solving the K\"ahler-BPS conditions
in which the holomorphic matrix ${\cal F}$ is derived from quantum-corrected 
holomorphic sections~\cite{dewitetal, antoetal, harveymoore, henningson,
kounnas}. 
While this procedure sounds right, one needs to have a careful thought
for its consistency.
In general perturbative renormalization around a black hole
background shows quite different structure from that around a flat spacetime.
In calculating perturbative corrections to physical quantities such as BPS
black hole mass and entropy, one thus needs first to evaluate quantum 
corrections to the holomorphic sections around the black hole 
background and then to use it for solving the
K\"ahler-BPS conditions. When deriving perturbative corrections to the
holomorphic sections, Refs.~\cite{dewitetal}--\cite{henningson} have
 assumed a flat spacetime, vanishing gauge field strengths and  
constant scalar fields\footnote{With a notable exception of Ref.~\cite{
kounnas}, where non-vanishing but constant background fields were 
considered.}.  
The K\"ahler-BPS black hole configuration is rather special in this aspect. 
It is distinguished from other BPS black holes by the fact that the scalar
fields are constant everywhere outside the horizon. Therefore, so long as
one restricts to a macroscopic K\"ahler-BPS black hole, whose horizon is 
large enough that spatial curvature and gauge field strengths at black hole
exterior are small enough, one may use the quantum-corrected holomorphic
sections of Ref.~\cite{harveymoore} when solving the Kahler-BPS 
conditions. In this section, we will adopt this strategy and {\sl a
posteriori} justify this procedure by showing that the scalar fields
remain constant everywhere, viz. the K\"ahler-BPS bound is saturated
at quantum level.  

\subsubsection{\rm PERTURBATIVE CORRECTIONS} 
Perturbative corrections to the heterotic prepotential has been calculated
by a direct threshold calculation and by mapping the classical prepotential
of type IIA string compactifications on Calabi-Yau three-folds~\cite{yaugroup}
 to the heterotic side using the type IIA-heterotic string duality~\cite{
kachruvafa}. Under the string duality, the type IIA worldsheeet instanton 
corrections to the type IIA prepotential is mapped to the spacetime instanton
corrections that depend on heterotic $S$-field and to the worldsheet instanton
corrections. The $N=2$ supersymmetry non-renormalization theorem guarantees 
that the quantum effects on the heterotic side comes from one-loop and
from non-perturbative effects. In this section, we study the 
perturbative one-loop correction exclusively. 
The non-perturbative corrections will be discussed in the next section. 
The non-renormalization theorem dictates that 
the perturbative corrections to the prepotential and to the 
holomorphic section depends only on $T, U$ moduli fields. 
In the large ${\rm Im}T, {\rm Im} U$ limit, 
the heterotic $STU$ model prepotential contains 
cubic polynomials in $T, U$, constant term and
exponentially suppressed ${\cal O}(e^{2 \pi i T}, e^{2 \pi i U})$ 
worldsheet instanton correction terms. The direct calculations~\cite{
harveymoore, dewitetal, antoetal} have shown that the target-space duality
symmetry $SL(2,{\bf Z})_T \times SL(2,{\bf Z})_U \times {\bf Z}_2^{
T \leftrightarrow U}$ acting on the moduli $T$ and $U$ constrains strongly 
these one-loop corrections.
Only third derivatives of $F^{(P)}$ with respect to 
$T$ and $U$ fields transform as well-defined modular forms under 
the modular group $SL(2,{\bf Z})_T \times SL(2, {\bf Z})_U$. 
Integrating back to obtain $F^{(P)}$ then poses a quadratic ambiguity 
that amounts to redefinition ambiguity of the dilaton 
$S \rightarrow S + \alpha T + \beta U$, where $\alpha, \beta$ are arbitrary
parameters. Comparing with the tree-level prepotential, one finds that 
this redefinition gives rise to an ambiguity to the perturbative part
of the prepotential $F^{(P)} \rightarrow F^{(P)} + \alpha T^2 U + \beta T U^2$. 
By adopting a convention\footnote{
Alternative possible choice is $\alpha = 0, \beta = -1$
~\cite{cardosochoice}
This choice ensures the $S \leftrightarrow T$ exchange symmetry~\cite{
stsymm} present in the theory. We are motivated to choose the above
convention as the K\"ahler-BPS configuration of the heterotic $S$-field
turns out to be invariant under the target-space duality perturbatively. 
Different choices, however, should be all physically equivalent.
}
 that the 
invariant dilaton is the same as the special coordinate dilaton, viz. $
\alpha = \beta = 0$, one finds that the perturbative correction to 
the prepotential is given by
\bee
F^{(P)}(T, U) = && (X^0)^2 (T^2 U + {1 \over 3} U^3) \,
\theta ({\rm Im} T - {\rm Im} U)
\nonumber \\
&+& (X^0)^2 (U^2T + { 1 \over 3} T^3) \,
\theta ({\rm Im} U - {\rm Im} T)
\nonumber \\
&+& 240 \cdot {\zeta(3) \over (2 \pi)^3}  
+ {\cal O} (e^{2i\pi T}, e^{2i \pi U}).
\label{quprepot}
\eee
The first and the second lines are the $S$-field redefinition ambiguity-free,
cubic polynomial part of the prepotential. It, however, depends on 
the Weyl chamber divided at ${\rm Im} T = {\rm Im} U$ that is
 symmetric under ${\bf Z}_2^{T \leftrightarrow U}$ exchange.
The third line denotes constant part, which depends on the famous
$\zeta(3)$ and the Euler number $\chi (CY) = - 480$ of the 
corresponding Calabi-Yau space and 
contributions of exponentially suppressed worldsheet instanton effects. 
It is important to note that, at quantum level,
  the heterotic $S$-field is a special
 coordinates but is not invariant under the target--space duality. 
As we will see, however, so long
as the nonperturbative quantum corrections are ignored, the K\"ahler-BPS
configuration of the $S$ field
remains target-space duality invariant. In fact, because of this, we have
chosen the quadratic ambiguity of the prepotential in the simplest manner, 
viz. the perturbative $S$-field is the same as the classical one. On the
other hand, it is {\sl not} the $S$ or $S_{\rm inv}$ that organizes
the perturbation expansions, but the one including threshold corrections. 
It is possible to define a renormalized
dilaton, which is also invariant under the target--space duality~\cite{
dewitetal,harveymoore}. In what
follows, since we are mainly in the weak coupling limit, we will start 
with the special coordinate dilaton, $S$. 

Holomorphic sections including the perturbative corrections are given
by 
\bee
X^\Lambda &=& (X^0, { X}^1, { X}^2, {X}^3) = 
(1 - TU, T, U),
\nonumber \\
{F}_\Lambda &=& F^{(0)}_\Lambda + F^{(P)}_\Lambda,
\nonumber \\
F^{(P)} &\equiv&
(-{1 \over 3} U^3 - T^2 U, \,\,0 \,\, , 2TU \, , T^2 + U^2)
.
\eee

Quantum correction to the holomorphic coupling matrix 
is easily calculated. The explicit change of it is given by
\be
{\cal F}^{(P)}_{\Lambda \Sigma} \equiv \partial_\Lambda F^{(P)}_\Sigma
= 
\left( \begin{array}{cccc}
{2 \over 3} U^3 + 2 T^2 U & \, 0 \, & \, -2TU \, & - T^2- U^2 \\
0 & \, 0 \,  & \, 0 \, & 0 \\
- 2TU & \, 0 \, & \, +2U \, & +2T \\
- T^2 - U^2 & \, 0 \, & \, +2T \, & +2 U \\
\end{array} \right).
\label{quanholmatrix}
\ee
One first notes that the entries in the second rows and columns are 
zero to any finite orders in perturbation theory.
Later, in identifying the perturbative quantum corrections to the
black hole BPS mass and the entropy, this observation will play
an important role.
One can also verify that the holomorphic matrix ${\cal F}^{(P)}$
satisfies the homogeneous degree-two condition:
$\partial_{ (\Delta} {\cal F}^{(P)}_{\Lambda \Sigma)} X^\Lambda = 0$.

The quantum corrected K\"ahler potential is easily derived:
\bee
K &=& - \log \Big[ {X}^\Lambda {\overline F}_\Lambda - 
{{\overline X}}^\Lambda {F}_\Lambda \Big]
\nonumber \\
& \equiv & [K_S + {\hat K} (T, U)]_{\rm Pert} \nonumber \\
K_S^{\rm Pert} &=& - \log[2 ({\rm Im} S + V_{GS}(T, U))] \nonumber \\
{\hat K}^{\rm Pert} &=& - \log [i  \langle X, {\overline X} \rangle ] =
 \log[2 {\rm Im} T \cdot 2 {\rm Im} U ].
\eee
One first notes that the K\"ahler potential of
the $T, U$ moduli fields is not modified at all by perturbative 
quantum effects.
This stems from the fact that the holomorphic sections $X^\Lambda$ 
does not contain the heterotic dilaton $S$. This in turn ensures that
the geometric nonlinear constraint 
\be
\langle X, X \rangle_{\rm Pert} = 0.
\label{quconst}
\ee
remains valid to all orders in perturbation theory.
All the one-loop quantum correction then goes to the so-called 
Green-Schwarz term $V_{GS}$ in the K\"ahler potential of the
heterotic $S$-field. 
This correction comes from universal threshold corrections for the 
heterotic rank-3 model:
\bee
V_{GS}(T, U) &=& i  {[2  - {\rm Im} T \partial_{{\rm Im} T} - {\rm Im} U 
\partial_{{\rm Im} U} ] ({F}^{(P)} + {\overline {F}}^{(P)} ) 
\over 4 {\rm Im} T {\rm Im } U} 
\nonumber \\
& = & - {\rm Im} T - {{1 \over 3} ({\rm Im} U)^2 \over {\rm Im T}}
+ 240 {\zeta(3) \over (2 \pi)^3} 
{ 1 \over ({\rm Im} T {\rm Im} U)}.
\eee

While the classical heterotic string coupling is governed by the
special coordinate $S$, 
at perturbative level, the heterotic string-loop expansion parameter is
a combination of dilaton and and Green-Schwarz term~\cite{dewitetal}:
\be
\Big[ {4 \pi \over g^2_{\rm het}} \Big]_{\rm Pert.}
={\rm Im} S + V^{\rm GS}
\equiv {\rm Im} S_{\rm inv} + V^{\rm GS}_{\rm inv}.
\label{hetcoupldef}
\ee

\subsubsection{\rm DYNAMICAL RELAXATION FOR QUANTUM BPS BLACK HOLE}

One now solves the K\"ahler-BPS conditions including the
perturbative quantum corrections
\be
{\rm M}_{\rm Pl} e^{K/2} \left( \begin{array}{c} {X}^\Lambda \\ {F}_\Lambda
\\ \end{array} \right)
= - { 1 \over 2 {\overline {\bf Z}}} \left( \begin{array}{cc}
i {\bf I} \hskip-0.16cm {\bf I} & ({\rm Im} {\cal F})^{-1} \\
- {\rm Im} {\cal F} & i {\bf I} \hskip-0.16cm {\bf I} \\ \end{array} \right)
\cdot {\bf Q}.
\label{qubpscond}
\ee
 
It turns out, as in the classical case, that the geometric
nonlinear constraint
Eq.(\ref{quconst}) plays an important role.
Inserting Eq.(\ref{qubpscond}) into Eq.(\ref{quconst}), one finds: 
\bee
&& \langle {\bf p} , {\bf p} \rangle_{\rm Pert}
 = \langle {\bf q} , {\bf q} \rangle_{\rm Pert}
\label{pertre} \\
&& ( {\bf p} \cdot {\bf q} )_{\rm Pert} = 0.
\label{pertim}
\eee
Note that, even though structure of the constraint looks the same as in 
the classical case, each components of the charge vector components
are expected to receive quantum corrections. Recall that ${\bf P} = P$ 
denotes the microscopic, integer-valued magnetic charges, hence, do not
change by quantum effects. The ${\bf q}$ charges, however, may be modified,
since it depends on the holomorphic matrix and scalar fields. For the STU
model, there are four components of ${\bf q}$. Imposing the two
constraints Eqs.(\ref{pertre},~\ref{pertim}) leaves two free components
of ${\bf q}$ that can adjust as quantum effects are included. The argument
clearly indicates nontrivial quantum effects to the K\"ahler-BPS 
configuration.

Quantum effects to the black hole mass and entropy arise in two possible ways.
One is through {\sl explicit} change of the K\"ahler potential
$K_S^{\rm Pert}$ by the Green-Schwarz term. Another is through {\sl
implicit} functional change of the scalar fields that depend on the 
charges ${\bf q}$ for K\"ahler-BPS configurations. 
Note that we have shown 
that components of ${\bf q}$ charges are not protected
at all from the quantum corrections.
 
\subsubsection{\rm PERTURBATIVE NON-RENORMALIZATION THEOREM OF BLACK-HOLE 
MASS \& ENTROPY}
We now establish perturbative non-renormalization of the K\"ahler-BPS 
black hole mass and entropy. Before doing so, we first show that there
is no implicit functional shift of the heterotic dilaton $S$-field from 
the classical configuration. From Eq.(\ref{qubpscond}), one finds quantum
corrections to the ${\rm Im} S$:
\be
\Delta ( {\rm Im} S) 
= \Delta \Big[ ({\rm Im} {\cal F})_{1 \Sigma}
\, {[i {\bf p} + {\bf q}]^\Sigma \over [i {\bf p} + {\bf q} ]^0 }
\Big].
\label{imspart}
\ee
On the right hand side, the quantum corrections arise both from an 
explicit change of the first factor by ${\rm Im} {\cal F}^{(P)}$
and from an implicit change of the ${\bf q}$ charge components.
One first notes that the first row and column of the perturbative
holomorphic matrix ${\cal F}^{(P)}$ in Eq.(\ref{quanholmatrix}) vanish
identically. While it was shown for large $T, U$ limit, this
holds true at finite $T, U$ as well\footnote{Note that the
constant and the infinite worldsheet instanton expansion terms 
also depends only on $T, U$ fields.} and merely reflects the
fact that the perturbative corrections are independent 
of the heterotic
dilaton $S$. The observation leads to a conclusion that there is no
explicit quantum correction from the first factor in Eq.(\ref{imspart}). 
Taking the classical
K\"ahler-BPS configuration for the first factor, one then also finds
that the charge ratios in the second factor cancel out, hence, no
implicit changes. A symplectic transformation that shifts the 
vacuum angle leads to the conclusion that 
${\rm Re} S$ is not renormalized either.
This completes the proof that the K\"ahler-BPS configuration of the 
heterotic dilaton $S$-field is not renormalized to all orders in 
perturbation theory.

We now establish the afore-mentioned non-renormalization theorem. Following
exactly the same steps as in the classical analysis in Eq.(\ref{bpsexp}),
consider the inner product of the perturbative sections $X^\Lambda$ that
satisfy Eq.(\ref{qubpscond}) with its complex-conjugate. This yields:
\be
4 {\rm M}_{\rm Pl}^2 \, [ e^K \, |{\bf Z}|^2
\, \langle X, {\overline X} \rangle \,]_{\rm Pert}
= [\langle {\bf p}, {\bf p} \rangle + \langle {\bf q},
{\bf q} \rangle ]_{\rm Pert},
\ee
hence,
\be
{\rm M}_{\rm Pl}^2 |{\bf Z}|^2_{\rm Pert}
= {1 \over 4} \Big( 
[e^{-K_S}] \, [e^{\hat K} \langle X, {\overline X} \rangle ]^{-1}
[\langle {\bf p}, {\bf p} \rangle + \langle {\bf q}, {\bf q} \rangle ] 
\Big)_{\rm Pert}  
\label{bpsmass1}
\ee
One now analyze quantum corrections to the right hand side 
of Eq.(\ref{bpsmass1}).
Since the form of the K\"ahler potential $\hat K$ 
and the holomorphic sections $X^\Lambda$ 
are
not changed perturbatively, the second factor is not renormalized and
remains unity. The third factor is not renormalized either since
\be
\Delta (\langle {\bf p}, {\bf p} \rangle + \langle
{\bf q}, {\bf q} \rangle)_{\rm Pert} 
= 2 \Delta \langle {\bf p}, {\bf p} \rangle_{\rm Pert}
= 2 \Delta \langle P, P \rangle = 0.
\label{zero}
\ee
In the first equality, the perturbative 
nonlinear constraint Eq.(\ref{pertre}) is used. The
second equality follows from the fact that ${\bf p}$ equals to the
microscopic, integer-valued magnetic charge $P$. This charge cannot 
jump by quantum effects. 
Since ${\rm Im} S$ is not renormalized as shown above, the only
perturbative correction to the Eq.(\ref{bpsmass1}) comes from the
explicit dependence through the Green-Schwarz term in 
$e^{-K_S}$. One thus concludes that:
\be
|{\bf Z}|^2_{\rm Pert}
= \exp [- (K_S^{\rm Pert.} - K_S^{\rm Cl})] \cdot |{\bf Z}|^2_{\rm Cl}
=
{( {\rm Im} S + {\rm V}^{\rm GS})_{\rm Pert}
 \over ({\rm Im} S)_{\rm Cl} } \cdot |{\bf Z}|^2_{\rm Cl}.
\label{ratio}
\ee

The perturbative BPS black hole mass and the entropy formula is then
obtained from Eq.(\ref{ratio}) as
\footnote{
Recently the authors of \cite{lustgroup} have made a conjecture for the
perturbatively corrected black hole mass/entropy formula. Our result disagrees
with theirs. That the formula we have obtained should be the correct one can
be understood simply in the following way.
It is known that the combination $( {\rm Im} S + V_{Gs})$
equals to the combination of so-called
invariant dilaton and the invariant Green-Schwarz term 
\be
{\rm Im} S + {\rm V}^{\rm GS} = {\rm Im} S_{\rm inv} + {\rm V}^{\rm
GS}_{\rm inv}.
\ee
This combination defines heterotic string-loop expansion parameter with
manifest target-space duality invariance.
Therefore it is natural to expect for
this combination to appear in  physical quantities such as BPS black hole
mass and entropy once evaluated in a consistent perturbative expansion.
} :
\bee
[ {\bf M}_{\rm BH}^2]_{\rm Pert} = \big( { S_{\rm BH} \over \pi} \Big)_{\rm
Pert} = {\rm M}_{\rm Pl}^2 |{\bf Z}|^2_{\rm Pert}
&=& \Big( {\rm Im } S + V^{\rm GS} ({\rm Im} T, {\rm Im} U) \Big)
\,\, \langle P, P \rangle
\nonumber \\
&=& \Big( {4 \pi \over g^2_{\rm het}} \Big)_{\rm Pert}
\,\, \langle P, P \rangle 
\label{pertmassentr}
\eee
where Eq.(\ref{hetcoupldef}) was used.
Note that, nowhere in deriving the non-renormalization theorem and 
Eq.(\ref{pertmassentr}), specific details of the perturbative correction 
to the prepotential were assumed except that the correction should
be independent of $S$. This is a property for any $N=2$ heterotic
string compactifications should satisfy, hence, the perturbative 
black hole mass and entropy formula Eq.(\ref{pertmassentr}) is expected
to be valid for any $N=2$ heterotic string compactifications\footnote{
This should also hold, in particular, for heterotic compactifications which
do not have known type-IIA side duals.} In particular, for the heterotic
STU model, the formula Eq.(\ref{pertmassentr}) is valid not only 
for large $T, U$ limit but also  for finite $T, U$ points 
so long as one stays away from the enhanced gauge symmetry points. 
Recall that the Green-Schwarz term originates from the universal threshold
corrections of heavy Kaluza-Klein and winding states, that are charged
under the $U(1) \times U(1)$ gauge group associated with the $T, U$
special coordinates. Since there are no massless charged fields coupled
to these gauge fields , the quantum correction is entirely summarized by 
the threshold corrections only.

It should be emphasized that it was the target-space duality symmetry
that guaranteed for the non-renormalization theorem to hold. Both the
fact that the K\"ahler potential ${\hat K}$ was not modified by perturbative
effect and the fact that the nonlinear constraint $\langle X, X \rangle
= 0$, from which the two important constraint Eqs.(\ref{pertre}, \ref{pertim})
were derived, were the two ingredients in deducing the non-renormalization
theorem from Eq.(\ref{bpsmass1}).

Actually, there exist finer structures to the non-renormalization theorem.
In deriving the perturbative central charge Eq.(\ref{bpsmass1}), we have
used the special aspect of $T_2$ compactification of the heterotic string.
In $X^0 = 1$ gauge, in general, the central charge was obtained from the
gauge fixing itself, see Eq.(\ref{bpscharge}). The same procedure with
perturbative corrections for the heterotic STU model yields:
\bee
{\rm M}_{\rm Pl}^2 |{\bf Z}|^2_{\rm Pert} 
&=& { 1 \over 4} e^{-K} \Big[
({\bf p}^0)^2 + ({\bf q}^0)^2 \Big] \nonumber \\
&=& { 1 \over 2} ({\rm Im} S + V^{GS}(T,U)) 
\cdot \Big[ e^{-{\hat K}(T,U)} \Big( ({\bf p}^0)^2 + ({\bf q}^0)^2 \Big) \Big].
\eee
Comparing this expression with Eq.(\ref{bpsmass1}), one finds a relation:
\be
[({\bf p})^2 + ({\bf q})^2] = e^{-{\hat K}} [ ({\bf p}^0)^2 + ({\bf q}^0)^2].
\label{relation}
\ee
In Eq.(\ref{zero}), it was argued that the left hand side is not renormalized.
Hence, on the right hand side of Eq.(\ref{relation}), 
renormalization of the two terms should 
cancel each other. The $T, U$ moduli fields, in general, receives nontrivial
renormalizations, as is evident from the ${\cal F}$ dependence of
Eq.(\ref{qubpscond}):
\bee
T &=& \Big({X^2 \over X^0} \Big)_{\rm Pert} = \Big( { [i {\bf p} + {\bf q}]^2
\over [ i {\bf p} + {\bf q}]^0} \Big)_{\rm Pert}, \nonumber \\
U &=& \Big( {X^3 \over X^0} \Big)_{\rm Pert}
= \Big( { [i {\bf p} + {\bf q} ]^3 \over [ i {\bf p} + {\bf q}]^0} \Big)_{\rm
Pert}.
\label{tufields}
\eee 
Such renormalizations of the $T, U$ moduli fields 
are string counterparts of those of the Higgs vacuum expectation values
in the rigid $N=2$ supersymmetric gauge theories. Despite the fact that
$T, U$ moduli fields are renormalized, we now show that the 
non-renormalization theorem is stronger enough that the two factors
on the right hand side of Eq.(\ref{relation}) are not renormalized 
separately. Again, it turns out the target-space duality symmetry is 
to ensure their non-renormalizations. 

\subsubsection{\rm PERTURBATIVE NON-RENORMALIZATION OF $[({\bf p}^0)^2 + 
({\bf q}^0)^2]$ }
Recall that the macroscopic charges are related to the
microscopic charges as:
\bee
{\bf p}^\Lambda &=& P^\Lambda \nonumber \\
{\bf q}^\Lambda &=& [({\rm Im} {\cal F})^{-1}]^{\Lambda \Sigma} \, 
[- {\rm Re} {\cal F} \cdot P + Q]_\Sigma.
\label{chargedefinition}
\eee
Since $P^\Lambda$'s are integer-valued, one finds:
\be
\Delta \Big([({\bf p}^0)^2 + ({\bf q}^0)^2 ] \Big) 
= \Delta ({\bf q}^0)^2.
\label{corr}
\ee 
The quantum corrections on the right hand side come in through 
explicit modification of the holomorphic matrix ${\cal F}$ and 
implicit modification of the moduli fields. Using the matrix identity
\be
{ 1 \over {\rm Im} {\cal F} } = { 1 \over {\rm Im} {\cal F}^{(0)}
+ {\rm Im} {\cal F}^{(P)} }
\equiv \sum_{n=0}^\infty { (-)^n \over {\rm Im} {\cal F}^{(0)} }
\Big( {\rm Im } {\cal F}^{(P)} { 1 \over {\rm Im} {\cal F}^{(0)} } \Big)^n,
\label{matrixexpansion}
\ee
the structure of the classical holomorphic matrix: 
\be
\Big[{1 \over {\rm Im} {\cal F}^{(0)} } \Big]^{\Lambda \Sigma}
= \Big( {1 \over {\rm Im} S } \Big) \eta^{\Lambda \Sigma},
\ee
and the fact the first row and column of the quantum correction part of
the holomorphic matrix ${\cal F}^{(P)}$ are zero, 
one finds that all higher-order $(n \ge 1)$ terms of $\Lambda = 0$ 
component in Eq.(\ref{matrixexpansion}) vanish identically.
Since $({\rm Re} {\cal F}^{(P)})_{1 \Sigma} = 0$ as well, one concludes that
\be
\Delta ({\bf q}^0) = \Delta \Big( ({1 \over {\rm Im} {\cal F} })^{0 \Sigma}
\cdot  [ - ({\rm Re} {\cal F}) \cdot P + Q ]_\Sigma \Big) = 0
\ee
In the previous subsection, the
${\rm Im} S$, which defines the $n=0$ classical term in ${\bf q}^0$, does
not receive quantum corrections. One thus concludes that $(({\bf p}^0)^2 
+ ({\bf q}^0)^2)$ is not renormalized to all orders in perturbation theory.

\subsubsection{ \rm PERTURBATIVE NON-RENORMALIZATION OF $\exp(-{\hat K})$}
We now examine $e^{-\hat K} = 4 {\rm Im} T \, {\rm Im} U$.
From Eq.(\ref{tufields}), one obtains
\bee
{\rm Im} T &= & { [ {\bf p}^2 {\bf q}^0 - {\bf p}^0 {\bf q}^2 ] 
\over [ ({\bf p}^0)^2 + ({\bf q}^0)^2] }  ,
\nonumber \\
{\rm Im} U &=& { [ {\bf p}^3 {\bf q}^0 - {\bf p}^0 {\bf q}^3 ] 
\over
[ ({\bf p}^0)^2 + ({\bf q}^0)^2 ] }.
\label{tusol}
\eee
The previous non-renormalization theorem guarantees that the denominators
are not renormalized. It now remains whether the numerators are 
renormalized. From the nonvanishing entries of ${\cal F}^{(P)}$, 
it should be clear that individual numerators receive 
quantum corrections. This is not surprising since ${\rm Im} T, {\rm Im} U$
are the string counterpart of the vacuum expectation values of the scalar
fields in the rigid theory. Surprising enough, we now show that the
renormalization effects cancel each other so that the product 
${\rm Im} T \cdot {\rm Im} U$, hence, the K\"ahler potential ${\hat K}$
is not renormalized at all. This is another manifestation of the 
target-space duality symmetry.

Consider the quantum correction to the product of the two numerators of
Eq.(\ref{tusol}):
\be
\Delta \Big( {\bf p}^2 {\bf p}^3 ({\bf q}^0)^2 - {\bf p}^ 0 {\bf p}^2 {\bf 
q}^0 {\bf q}^3 - {\bf p}^0 {\bf p}^3 {\bf q}^0 {\bf q}^2 
+ ({\bf p}^0 )^2 {\bf q}^2 {\bf q}^3 \Big) .
\label{numerator}
\ee
Since ${\bf p}^\Lambda = P^\Lambda$ are moduli-independent, integer-valued 
and ${\bf q}^0$ charges are not renormalized perturbatively, we focus only
on ${\bf q}^{1,2,3}$ charges. The constraint $\langle X, X \rangle_{\rm Pert}
 = 0$ puts two conditions:
\bee
{\bf q}^0 \Delta {\bf q}^1 + \Delta ({\bf q}^2 {\bf q}^3) &=& 0, 
\nonumber \\
{\bf p}^0 \Delta {\bf q}^1 + {\bf p}^2 \Delta {\bf q}^3 + {\bf p}^3 
\Delta {\bf q}^2 &=& 0.
\label{constraintrela}
\eee
Using the two constraints, it is straightforward to check that
 Eq.(\ref{numerator}) vanishes identically.
This establishes the perturbative non-renormalization theorem of the K\"ahler
potential ${\hat K}$, a product of two Higgs expectation values, as protected
by the target-space duality\footnote{ As we have shown above, the individual 
$T$ and $U$ fields are subject to renormalization. 
It is important to note that a specific form of the radiative correction
to these fields are meaningless unless one specifies how the cubic 
polynomial ambiguity in the one-loop corrected prepotential is fixed. 
In the rigid $N=2$ supersymmetric theories, similarly, 
the Higgs expectation value by itself is not a meaningful quantity 
unless renormalization prescription of the gauge coupling constant is
specified.}.   

\subsection{Nonperturbative Quantum Corrections}
So far, we have ignored possible nonperturbative effects such as
corrections due to spacetime gauge or gravitational instantons. 
Such instanton effects give rise to holomorphic but exponentially suppressed
corrections to the holomorphic sections. At finite ${\rm Im} S$,
such non-perturbative effects are intractibly complicated.
In this section, we limit ourselves to the weak coupling limit,
\be
{\rm Im} S \gg {\rm Im} T \gg {\rm Im} U \gg 1
\label{weakcoupling}
\ee
and calculates the leading--order non--perturbative corrections. 

The classical Peccei-Quinn symmetry constrains the nonperturbative 
correction to the prepotential to be of a form:
\be
{\cal F} = (X^0)^2 [ STU + F^{(P)} (T, U)
+ F^{(NP)} (e^{ 2 \pi i S}, T, U) ].
\label{nonpprep}
\ee
After the strong-weak coupling symplectic transformation, one finds that
the holomorphic sections are further corrected:
\bee
X^{\Lambda, ( NP)} &=& ( 0, - \partial_S F^{(NP)}, 0, 0 )
\nonumber \\
F_\Lambda^{(NP)} &=& 
( (2 - S \partial_S - T \partial_T - U \partial_U) F^{(\rm NP)}, 0, 
\partial_T {F}^{(\rm NP)}  , \partial_U {F}^{(\rm NP)} ),
\eee
where the superscript NP denotes nonperturbative correction parts to the
 sections. 
In the weak coupling limit Eq.(\ref{weakcoupling}), one finds the 
non-perturbative K\"ahler potential as:
\bee
K_{\rm Nonp}  &\rightarrow&  K_S + [{\hat K}]_{\rm Pert} + 
\Delta {\hat K }_{\rm NP}
\nonumber \\
\Delta {\hat K}_{\rm NP} &=& {  {\rm Im} S \over {\rm Im} T {\rm Im} U}
\cdot {\rm Im} ( \overline{TU} \partial_S F^{\rm (NP)} )
= {\cal O} \Big( {\rm Im} S \, e^{- 2 \pi {\rm Im} S  } \Big)
\label{nonpkahler}
\eee
where $\Delta {\hat K}^{\rm (NP)}$ 
denotes the leading-order non-perturbative corrections
in the limit Eq.(\ref{weakcoupling}).
Here, we have used the fact that $\partial_{S,T,U} F^{(NP)} = {\cal O} ( 
F^{(NP)})$ and Eq.(\ref{weakcoupling}).
Note that the non-perturbative correction $\Delta {\hat K}^{\rm (NP)}$ 
is proportional to ${\rm Im} S$, the classical heterotic coupling constant.

One observes
two important changes compared to the pattern of the perturbative 
corrections. 
First, one finds that the holomorphic matrix ${\cal F}$ receives further
nonperturbative corrections. The first row and column entries, which
was vanishing perturbatively, are now nonzero:
\be
{\cal F}^{(NP)} = 
\left( \begin{array}{cccc}
({\rm Pert}) & \partial_S {\cal D} 
F^{(\rm NP)} & ({\rm Pert}) & 
({\rm Pert}) \\
\partial_S {\cal D} F^{\rm (NP)} & 0 & \partial_S \partial_T F^{(NP)}
& \partial_S \partial_U F^{(NP)} \\
({\rm Pert}) & \partial_T \partial_S F^{\rm (NP)} & ({\rm Pert}) & 
({\rm Pert}) \\
({\rm Pert}) & \partial_U \partial_S F^{\rm (NP)} & ({\rm Pert}) &
({\rm Pert}) \\
\end{array} \right)
\ee
where ${\cal D} \equiv
(2 - S \partial_S - T \partial_T - U \partial_U) $ 
and (Pert) denotes the entries that were non-vanishing already at the 
perturbative level. Recalling the proof of perturbative non-renormalization
theorems, it is clear that the non-vanishing first row and column entries 
render the $S$-field is renormalized non-perturbatively. As in the
rigid supersymmetric field theories, the correction may 
be interpreted as non-perturbative renormalization of the heterotic 
string loop-counting parameter.
Second,
it is not only $F_\Lambda$'s but also $X^\Lambda$'s that receive corrections.
In particular, the nonlinear constraint $\langle X, X \rangle  =  2 \partial_S
F \ne 0$. This indicates 
that the 
$T, U$ moduli fields should be further renormalized in addition 
to the perturbative renormalization. 

In the weak coupling limit Eq.(\ref{weakcoupling}) the  
non-perturbative violation of the geometric constraint $\langle X, X \rangle
= 0$ is exponentially small. Therefore, making an expansion in powers of
${\rm Im} S$ and $\exp (-{2 \pi {\rm Im} S})$,  
the leading non-perturbative corrections to the BPS central charge is 
given by 
\bee
 {\rm M}_{\rm Pl}^2 |{\bf Z}|^2_{\rm NP} 
&=& {1 \over 4} \Big( e^{-K} \, [({\bf p}^0)^2 + ({\bf q}^0)^2]
\Big)_{\rm Nonp}
\nonumber \\
&=&
{1 \over 2} \Big( ({\rm Im} S + V^{\rm GS})_{\rm Nonp} \,
e^{- \Delta {\hat K}^{\rm NP}} 
\Big)
\cdot \Big( {\rm Im} T {\rm Im} U [({\bf p}^0)^2 + ({\bf q}^0)^2]
\Big)_{\rm Nonp}
\eee
up to ${\cal O}(e^{- 2 \pi {\rm Im} S})$.
In the last line, we have re-organized the leading order non-perturbative
corrections such that the second bracket corresponds to the combination
that were subject to the perturbative non-renormalization theorem.
Even in this limit, however, it is not clear if the perturbative 
non-renormalization theorems established for each factors inside the second
bracket remain uuchanged. In fact, as will be shown below, 
each non-renormalization theorems are violated non-perturbatively. 
However, we find 
the second bracket as a whole is subject to a new non-renormalization 
theorem.  

\subsubsection{\rm NON-PERTURBATIVE RENORMALIZATION OF HETEROTIC 
COUPLING CONSTANT}
We first analyze how the heterotic string-loop counting parameter is 
renormalized non-perturbatively. The minimization condition equation
\be
{\rm Im} S = ({\rm Im} {\cal F})_{1 \Sigma} 
{ [i {\bf p} + {\bf q} ]^\Sigma \over
[i {\bf p} + {\bf q} ]^0}
\ee
suggests that the nonperturbative corrections may arise through either
the overall holomorphic matrix or through the ${\bf q}^0$ charge vectors.
The definition of ${\bf q}^0$ in Eq.(\ref{chargedefinition}) 
shows clearly that it is renormalized non-perturbatively.
Recall that the perturbative non-renormalization of ${\bf q}^0$ 
was solely based on the fact that the first row and column entries of 
the ${\cal F}^{(P)}$ matrix was zero identically. 
For the $S$-field, however, the effect of nonperturbative renormalization
of ${\bf q}^0$ is cancelled out in the ratio, and the nonperturbative 
correction comes entirely from the $({\rm Im} {\cal F})_{1 \Sigma}$ parts.
Expanding in terms of components and using the minimal configuration of
$T, U$ special coordinates, one finds leading non-perturbative correction as
\bee
\Delta [    {\rm Im} S] &=& \Big( \partial_S {\cal D}
F^{(NP)} + T \partial_T F^{(NP)} + U \partial_U F^{(NP)}  \Big) 
\nonumber \\
&=& {\rm Im} \Big( \partial_S (2 - S \partial_S) F^{(NP)}  \Big)
\equiv V^{(\rm NP)} = {\cal O}({\rm Im} S \, e^{- 2 \pi {\rm Im} S}).
\eee
This correction is proportional to ${\rm Im} S$, hence, 
is the same order as the 
correction to the K\"ahler potential Eq.(\ref{nonpkahler}).
We are thus motivated to define a non-perturbative heterotic 
string coupling parameter as:
\be
\Big({1 \over g^2_{\rm het}   } \Big)_{\rm Nonp}
\equiv  [(  {\rm Im} S + {\rm V}^{\rm GS})_{\rm Cl} + V^{\rm (NP)} ] \cdot
e^{-\Delta {\hat K}_{\rm NP}     } .
\label{nonpcoupl}
\ee
The leading order non-perturbative corrections give rise to non-perturbative
renormalization of the heterotic coupling parameter.

\subsubsection{\rm NON-PERTURBATIVE NON-RENORMALIZATION OF ${\rm Im} T {\rm Im}
U [({\bf p}^0)^2 + ({\bf q})^2]$ }
That the charge-squared sum and the K\"ahler potential 
${\hat K}_{\rm Pert} = {\rm Im} T {\rm Im} U$ 
are non-perturbatively renormalized is again 
from the fact that the first row
and column entries of the holomorphic matrix ${\cal F}^{(NP)}$ are 
non-perturbatively corrected.
This in turn implies, as we have shown above, that ${\bf q}^0$ charge is 
renormalized and that the $T, U$ fields receives further corrections.
Nevertheless, we now establish a non-perturbative non-renormalization 
theorem that
\be
{\tt M}^2 \equiv {\rm Im} T {\rm Im} U \, [({\bf p}^0)^2 + ({\bf q}^0)^2]
\ee
is not renormalized non-perturbatively. In establishing the theorem, one 
again finds the constraint $\langle X, X \rangle = 0$, valid up to the
order ${\cal O}( e^{- 2 \pi {\rm Im } S} )$,  plays a crucial
role\footnote{Recall that the leading-order corrections to the 
non-perturbative heterotic coupling parameter was ${\cal O}({\rm Im} S \,
e^{- 2 \pi {\rm Im} S})$, hence, larger than the violation of the geometric
constraint.}. 
Noting that ${\bf q}^0$ is non-perturbatively renormalized, the
real and imaginary part of this constraint now reads:
\bee
{\bf p}^0 \Delta {\bf q}^1 + {\bf p}^1 \Delta {\bf q}^0 + {\bf p}^2 
\Delta {\bf q}^3 + {\bf p}^3 \Delta {\bf q}^2 &=& 0,
\nonumber \\
\Delta ({\bf q}^0 {\bf q}^1 + {\bf q}^2 {\bf q}^3) = 
\Delta ({\bf p}^0 {\bf p}^1 + {\bf p}^2 {\bf p}^3 ) &=& 0.
\label{nonpconst}
\eee

Consider expanding the nonperturbative correction
\bee
\Delta {\tt M}^2 &=& 
\Delta \Big( ({\rm Im} T) ({\rm Im} U) [({\bf p}^0)^2 + ({\bf q}^0)^2] \Big) 
\nonumber \\
&=& [({\bf p}^0)^2 + ({\bf q}^0)^2]^{-2} \nonumber \\
& \times & \Big[ \,\, ({\bf p}^0)^2 \Big( 
 \,\,\, \{   {\bf p}^2 {\bf p}^3 \Delta ({\bf q}^0)^2
   + ({\bf p}^0)^2 \Delta ({\bf q}^2 {\bf q}^3 ) \}
\nonumber \\
&& \hskip1.25cm 
- \{ {\bf p}^0 {\bf p}^2 \Delta ({\bf q}^0 {\bf q}^3) 
   - {\bf p}^0 {\bf p}^3 \Delta ({\bf q}^0 {\bf q}^2 ) \} \,\,\, \Big)
\nonumber \\
&&+ {\bf p}^0 {\bf p}^2 \Big( {\bf q}^0 {\bf q}^3 \Delta ({\bf q}^0)^2 
                             - ({\bf q}^0)^2 \Delta ({\bf q}^0 {\bf q}^3) \Big)
\nonumber \\
&&+ {\bf p}^0 {\bf p}^3 \Big( {\bf q}^0 {\bf q}^2 \Delta({\bf q}^0)^2 
                             - ({\bf q}^0)^2 \Delta ({\bf q}^0 {\bf q}^2) \Big)
\nonumber \\
&&- ({\bf p}^0)^2 \Big( {\bf q}^2 {\bf q}^3 \Delta ({\bf q}^0)^2
                       - ({\bf q}^0)^2 \Delta ({\bf q}^2 {\bf q}^3) \Big)
\, \, \Big].
\eee
Using the first relation of Eq.(\ref{nonpconst}), one can simplify this to
\bee
\Delta {\tt M}^2 &=&
[({\bf p}^0)^2 + ({\bf q}^0)^2 ]^{-2} \nonumber \\
&\times& \Big( \,\,
 ({\bf p}^0)^2 \Big[ ({\bf p}^0 {\bf p}^1 + {\bf p}^2 {\bf p}^3 )
\Delta ({\bf q}^0)^2 + ({\bf p}^0)^2 \Delta ({\bf q}^0 {\bf q}^1 
+ {\bf q}^2 {\bf q}^3) \Big] \nonumber \\
&& + 
({\bf p}^0)^2 \Big[ ({\bf q}^0)^2 \Delta ({\bf q}^0 {\bf q}^1 + {\bf q}^2 
{\bf q}^3 )
- ({\bf q}^0 {\bf q}^1 + {\bf q}^2 {\bf q}^3) \Delta ({\bf q}^0)^2 \Big]
\,\, \Big).
\eee
Using the second relation of Eq.(\ref{nonpconst}), one finds 
that this expression vanishes identically.
\subsubsection{\rm NON-PERTURBATIVE BPS BLACK HOLE MASS AND ENTROPY}
Based on the results of the preceding two subsections, we now have
non-perturbative result of the BPS central charge 
in the weak coupling limit Eq.(\ref{weakcoupling}):
\bee
|{\bf Z}|^2_{\rm Nonp} &=& \exp [- (K_S^{\rm Nonp} - K_S^{\rm Cl} )]
\cdot |{\bf Z}|^2_{\rm Cl} 
\nonumber \\
&=& {
[  (  {\rm Im} S + {\rm V}^{\rm GS} )_{\rm Cl} + V^{\rm (NP)} ] 
\cdot e^{-\Delta {\hat K}_{\rm NP} }  \over ({\rm Im} S)_{\rm Cl} 
} 
\, |{\bf Z}|^2_{\rm Cl}.
\eee
Therefore, using the definition of heterotic string coupling parameter
Eq.(\ref{nonpcoupl}), the BPS black hole mass and the entropy 
formula with the leading order non-perturbative correction is obtained as:
\bee
\Big[{\bf M}_{\rm BH}^2 \Big]_{\rm Nonp}
= M_{\rm Pl}^2 |{\bf Z}|^2_{\rm Nonp}
&=& 
[({\rm Im} S + {\rm V}^{\rm GS})_{\rm Cl} + V^{\rm (NP)} ] \,
e^{-\Delta {\hat K}_{\rm NP}} \,  \cdot \,  \langle P, P \rangle
\nonumber \\
&\equiv& \Big[ { 4 \pi \over g^2_{\rm het} } \Big]_{\rm Nonp} \,\,\cdot \,\, 
\langle P, P \rangle.
\eee

\section{DISCUSSION}
\setcounter{equation}{0}

In this paper, we have studied classical and quantum configurations of
the K\"ahler-BPS black holes in $D=4, N=2$ heterotic string compactification. 
We have first given an interpretation of the K\"ahler-BPS limit as a 
dynamical relaxation phenomena of the scalar fields. 
We have emphasized that the interpretation offers a unified description
of the K\"ahler-BPS limit for both rigid and local $N=2$ supersymmetric
cases.
We have 
analyzed quantum corrections to the BPS black hole mass and macroscopic
entropy. Much as in the rigid supersymmetric theories, we have shown a 
non-renormalization theorem that protects the saturation of the BPS bound 
at perturbative level: BPS black hole mass and entropy 
have exactly the same form as the classical one once re-expressed in 
terms of renormalized parameters.
The non-renormalization theorem ensures, among others, that strong-coupling
extrapolation of microscopic state counting for suitable D-brane 
configurations can be made to the macroscopic black hole mass and entropy
configurations once short-distance quantum effects are correctly 
identified and included.
We have shown that the perturbative non-renormalization theorem is valid
not only for large $T_2$ limit but also everywhere inside the $T, U$ moduli 
space so long as one stays away from the enhanced gauge symmetry points.
The non-perturbative corrections are more involved. We have only estimated
the leading-order corrections in the weak coupling, large $T_2$ limit.
Clearly, a better understanding of non-perturbative effects is desirable. 

We finally discuss the quantum heterotic BPS black holes from different
string theory point of view that are related to the heterotic STU model
by string-string duality.   
As mentioned already, the heterotic STU model is related to sequential 
$T_2$ compactification of the $D=6$ type I orientifold model on 
$K_3$ orbifold $T^4 / {\bf Z}_2$~\cite{seiberggroup}. 
At $D=6$ the type I dilaton belongs to a hyper multiplet, while the
Kaluza-Klein volume of the $K_3$ orbifold volume belongs to the tensor
multiplet and controls the strength of the gauge interactions.
While we have studied the rank-3 STU model as a concrete example,
it is straightforward to generalize to higher-rank models.  
Of particularly interesting situation in the heterotic side is the 
non-perturbatively generated vector multiplets which originates
from the non-perturbatively generated $D=6$ tensor multiplets 
due to small instantons~\cite{smallinstanton}. 
At $D=4$ the gauge coupling corresponding these non-perturbatively
generated vector-tensor multiplets is proportional to $T$-moduli field.
As such, BPS black hole carrying these vector-tensor
fields would be intrinsically quantum-mechanical.  
On the type I side, these vector-tensor multiplets are mapped to
the vector fields that arise from the new open string sector associated
with 5-branes in $D=6$. Their interaction strength is determined by
the volume of $K_3$ volume, hence, independent of type I dilaton.
Thus, one expects that BPS black holes on the type I side provide a  
clean description of the intrinsically quantum-mechanical BPS black
holes in the heterotic side. 

Alternatively, one may use the heterotic-type II duality~\cite{kachruvafa}
and map the heterotic $N=2$ BPS black hole to those on dual IIA string 
Calabi-Yau compactification. The prepotential of type IIA side encodes
topological data of the Calabi-Yau three-fold under consideration.
For example, the type IIA prepotential contains intersection numbers
$d_{ABC} = \int J_A \wedge J_B \wedge J_C$, 
Euler number $\chi(CY)$ and infinitely many 
rational instanton numbers.
The heterotic-type IIA duality maps these data to the heterotic 
couplings in a well-defined manner. 
The type II dilaton belongs to the hyper multiplets, hence, the 
prepotential arises classically only.
Since, as we have shown, the $N=2$ black hole configuration depends 
on details of the prepotential, the type IIA topological data are 
then map to various physical quantities associated with the 
BPS black hole such as mass and entropy. We have shown that, in the 
heterotic weak coupling limit, these quantities are protected by 
non-renormalization theorems. 
The corresponding BPS black holes on type IIA side should then be 
protected by a worldsheet non-renormalization theorem.  
Furthermore, there are other topological data of Calabi-Yau three-folds
on type IIA side. For example, gravitational curvature-squared coupling 
encodes informations of second Chern class $c_2 (J_A) 
\equiv \int c_2 \wedge J_A$ and elliptic instanton numbers, while
higher-order gravitational or gauge-gravitational couplings encode
higher-genus instanton numbers of the Calabi-Yau three-fold. 
Again, they have precise map on the heterotic side as higer-order
gauge and gravitational interaction couplings~\cite{dewitlatest}.
These couplings are also strongly constrained by the target-space duality
symmetry, hence, the non-renormalization theorem we have proven in this
work should be applicable to these couplings. 
Of particular interest on the heterotic side is the effect of 
the gravitational curvature-squared term since it appears already at 
the classical level\footnote{Higher-order gravitational couplings arise only
at quantum level.}. Thus, already at heterotic classical level,   
the BPS black hole should carry information of the second Chern class 
$c_2 (J_A)$ in addition to the classical intersection numbers of the
corresponding Calabi-Yau three-fold on the type IIA side.
Explicit constructions of
heterotic BPS black holes and a generalization of the
non-renormalization theorem will be reported
elsewhere.  Thus, plethora of topological data of the Calabi-Yau three-fold
encoded through holomorphic couplings suggest that 
it would be extremely interesting to find 
other macroscopic physical quantities BPS black holes than the mass
and entropy that are stable against extrapolation to the strong coupling
regime. Like mass and entropy, such quantities will also depend on
the topological data of Calabi-Yau space in a definite manner. 
Therefore, if one finds sufficiently many such physical quantities,  
it should then be possible to glean various topological data of Calabi-Yau 
space out of macroscopic black hole configurations. 

I thank to S.R. Das, M.K. Gaillard, R. Kallosh, 
D. Lowe, S. Mathur, L. Susskind, 
S. Theisen and S. Yankielowicz for discussions on subjects related to 
this work.
While preparing for this manuscript, we have received the preprint~\cite{
lustkallosh}, which contains overlapping independent result with part of the
present work.



\begin{thebibliography}{99}

\newcommand{\NP}[3]{{ Nucl. Phys.} {\bf #1} {(19#2)} {#3}}
\newcommand{\PL}[3]{{ Phys. Lett.} {\bf #1} {(19#2)} {#3}}
\newcommand{\PRL}[3]{{ Phys. Rev. Lett.} {\bf #1} {(19#2)} {#3}}
\newcommand{\PR}[3]{{ Phys. Rev.} {\bf #1} {(19#2)} {#3}}
\newcommand{\IJ}[3]{{ Int. J. Mod. Phys.} {\bf #1} {(19#2)} {#3}}
\newcommand{\CMP}[3]{{ Comm. Math. Phys.} {\bf #1} {(19#2)} {#3}}
\newcommand{\PRp} [3]{{ Phys. Rep.} {\bf #1} {(19#2)} {#3}}



\bibitem{stromingervafa} A. Strominger and C. Vafa, Phys. Lett.
{\bf 379B} (1996) 99.

\bibitem{horowitzstrominger} 

G.T. Horowitz and A. Strominger, Phys. Rev. Lett. {\bf 77} (1996) 2368;\\
G.T. Horotiwz, J. Maldacena and A. Strominger, Phys. Lett. {\bf 383B}
(1996) 151. 

\bibitem{maldacenacallan} C.G. Callan and J. Maldacena, 
Nucl. Phys. {\bf B472} (1996) 591.

\bibitem{myeretal} J. Maldacena and A. Strominger,
Phys. Rev. Lett. {\bf 77} (1996) 428; \\
C. Johnson, R. Khuri and R. Myers, Phys. Lett. {\bf B378} (1996) 78.

\bibitem{hawking} J.D. Bekenstein, Phys. Rev. {\bf D12} (1975) 3077;\\
S. Hawking, Phys. Rev. {\bf D13} (1976) 191.

\bibitem{sduality} A.~Font, L.~Ib\'a\~nez, D.~L\"ust and F.~Quevedo,
  Phys. Lett. {\bf B 249} (1990) 35;\\
  S.-J.~Rey, Phys. Rev. {\bf D 43} (1991) 256;\\
  A.~Sen, Phys. Lett. {\bf B 303} (1993) 22, {\bf B 329} (1994) 217;\\
  J.~Schwarz and A.~Sen,  Nucl. Phys.  {\bf B 411} (1994) 35; \\
  A. Sen, \IJ{A 9}{94}{3707}.

\bibitem{hulltownsend}
C. M. Hull and P. Townsend, \NP{B 438}{95}{109}.

\bibitem{witten} E. Witten,  \NP{B 443}{95}{85}.

\bibitem{kachruvafa} 
S. Kachru and C. Vafa, Nucl. Phys. {\bf B 450} (1995) 69.

\bibitem{ferraravafa}
S. Ferrara, J.A. Harvey, A. Strominger and C. Vafa, Phys. Lett. {\bf 361B}
(1995) 59.

\bibitem{polchinskiwitten}
J. Polchinski and E. Witten,
Nucl. Phys. {\bf B460} (1996) 525.

\bibitem{horavawitten}
P. Horava and E. Witten,
Nucl. Phys. {\bf B460} (1996) 506; {\sl ibid.} {\bf B475} (1996) 94. 

\bibitem{polchinski}
J. Polchinski, Phys. Rev. Lett. {\bf 75} (1995) 4274.

\bibitem{wilczek}
F. Larsen and F. Wilczek, Phys. Lett. {\bf 375B} (1996) 37.

\bibitem{cvetictseytlin}
M. Cvetic and A. Tseytlin, Phys. Rev. {\bf D53} (1996) 5619.

\bibitem{ferrarakalloshstrominger}
S. Ferrara, R. Kallosh and A. Strominger, Phys. Rev. {\bf D52 }\rm (1995) 5412.

\bibitem{strominger}
A. Strominger, Phys. Lett. {|bf 383B} (1996) 39.

\bibitem{ferrarakallosh}
S. Ferrara and R. Kallosh, 
Phys. Rev. {\bf D54 } (1996) 1514; 
{\sl ibid.} {\bf D54} (1996) 1525.

\bibitem{kalloshstudent1}
R. Kallosh, M. Shmakova and W.-K. Wong, 
{\tt hep-th/9607077}.

\bibitem{kalloshstudent2}
K. Behrndt, R. Kallosh, J. Rahmfeld, M. Shmakova and W.-K. Wong,
{\tt hep-th/9608059}.

\bibitem{lustgroup} G.L. Cardoso, D. L\"ust and T. Mohaupt,
{\tt hep-th/9608099}. 

\bibitem{specialgeometry}
A. Strominger, Comm. Math. Phys. {\bf 133} \rm (1990) 163;
\\
L. Castellani, R. D'Auria and S. Ferrara, Phys. Lett. {\bf 241B} (1990) 57 
and Class. Quant. Grav. {\bf 7} \rm (1990) 1767.

\bibitem{strominger2} 
D.M. Kaplan, D.A. Lowe, J.M. Maldacena and A. Strominger,
{\tt hep-th/9609024}.

\bibitem{seibergwitten}
N. Seiberg and E. Witten, Nucl. Phys. {\bf B426} (1994) 19;
{\bf 430} (1994) 485(E);
{\sl ibid.} {\bf B431} (1994) 484.

\bibitem{berglund} P. Berglund, S. Katz, A. Klemm and P. Mayr, 
{\tt hep-th/9605154}.

\bibitem{theisen} J. Louis, J. Sonnenschein, S. Theisen and S. 
Yankielowicz, {\tt hep-th/9606049}.

\bibitem{furthertest}
V. Kaplunovsky, J. Louis and S. Theisen, Phys. Lett. {\bf 357B}
(1995) 71;\\
I. Antoniadis, E. Gava, K.S. Narain and T.R. Taylor, Nucl.
Phys. {\bf B455} (1995) 109;\\
S. Kachru, A. Klemm, W. Lerche, P. Mayr and C. Vafa, Nucl. Phys.
{\bf B459} (1996) 537;\\
I. Antoniadis and H. Partouche, Nucl. Phys. {\bf B460} (1996) 475;
\\
G.L. Cardoso, G. Curio, D. L\"ust, T. Mohaupt and S.-J. Rey,
Nucl. Phys. {\bf B464} (1996) 
18.


\bibitem{seiberggroup} N. Seiberg and E. Witten, 
Nucl. Phys. {\bf B471} (1996) 121;\\
M. Berkooz, R. Leigh, J. Polchinski, J.H. Schwarz, N. Seiberg
and E. Witten, Nucl. Phys. {\bf B475} (1996) 115; \\
I. Antoniadis, C. Bachas, C. Fabre, H. Partouche and T.R. Taylor,
{\tt hep-th/9608012}.



\bibitem{jackson}
See, for example, Chap. 1.4, 4 and 7 of 
A.D. Jackson, {\sl Classical Electrodynamics}, 2nd Ed., 
John Wiley \& Sons, (New York, 1975).

\bibitem{inequality} E. Beckenbach and R. Bellman, {\em Inequalities} , 
Springer-Verlag (Berlin, 1975).

\bibitem{witteneffect} E. Witten, Phys. Lett. {\bf 86B} (1979) 283.

\bibitem{dynamicalrelax}
E. Cremmer, S. Ferrara, C. Kounnas and D. Nanopoulos, Phys. Lett. 
{\bf 133B} (1983) 61;
\\
J. Ellis, K. Enqvist and D. Nanopoulos, Phys. Lett. {\bf 151B} (1985) 357;
\\
S. Ferrara, C. Kounnas and F. Zwirner, Nucl. Phys. {\bf B429} (1994) 589 and
 {\bf B433} 255.

\bibitem{wittenposenergy} E. Witten, Comm. Math. Phys. {\bf 80} (1981) 381;
\\
J.M. Nester, Phys. Lett. {\bf 83A} (1981) 241.

\bibitem{gibbonshull} G. Gibbons and C. Hull, Phys. Lett. {\bf 109B} 
(1982) 190.

\bibitem{topfreeenergy}
S. Ferrara, C. Kounnas, D. L\"ust and F. Zwirner, Nucl. Phys. {\bf B365}
(1991) 431; 
\\
H. Ooguri and C. Vafa, Mod. Phys. Lett. {\bf A5} (1990) 1389.

\bibitem{narain}
K.S. Narain, Phys. Lett. {\bf 169B} (1986) 41;
\\
K.S. Narain, M.H. Sarmadi and E. Witten, Nucl. Phys. {\bf B278} (1987) 369.

\bibitem{ferraraetal} S. Ferrara, L. Girardello, C. Kounnas and M. Porrati,
Phys. Lett. {\bf 912B} (1987) 368;
\\
P. Fr\'e and P. Soriani, Nucl. Phys. {\bf B371} (1992) 659;
\\
A. Ceresol\'e, R. D'Auria and S. Ferrara, Phys. Lett. {\bf 339B} (1994) 71;
\\
A. Ceresol\'e, R. D'Auria, S. Ferrara and A. van 
Proeyen, Nucl. Phys. {\bf B444} (1995) 92.


\bibitem{ibanez} G. Aldazabal, A. Font, L. Ib\'a\~nez and F. Quevedo,
Nucl. Phys. {\bf B461} (1996) 85; {\tt hep-th/9602097}.

\bibitem{n2nonren} R. Barbieri, S. Ferrara, L. Maiani, F. Palumbo and
C.A. Savoy, Phys. Lett. {\bf 115B} \rm (1982) 212. 

\bibitem{barbieri}
R. Barbieri and S. Cecotti, Z. Phys. {\bf C17} (1983) 183; 
\\
M. Grisaru, A. Santambrogio and D. Zanon, {\tt hep-th/9610001}.

\bibitem{gaillard}
M.K. Gaillard and V. Jain, Phys. Rev. {\bf D46} (1992) 1786; Phys. Rev.
{\bf D49} (1994) 1951;
\\
M.K. Gaillard, V. Jain and K. Saririan, {\tt hep-th/9606052},
{\tt hep-th/9606135}.

\bibitem{wittenolive} E. Witten and D. Olive, Phys. Lett. {\bf 78B}
(1978) 97.

\bibitem{divecchia} A. D'Adda, R. Horsley and P. Di Vecchia, Phys. Lett.
{\bf 76B} \rm (1978) 298.

\bibitem{kaul} C. Imbimbo and S. Mukhi, Nucl. Phys. {\bf B249} \rm
(1985) 143 ; \\
R. Kaul, Phys. Lett. {\bf 143B} \rm (1984) 427.


\bibitem{dewitetal} 
B. de Wit, V. Kaplunovsky, J. Louis and D. Lust, Nucl. Phys. {\bf B451}
(1996) 53.

\bibitem{antoetal} I. Antoniadis, S. Ferrara, E. Gava, K. Narain and
T. Taylor, Nucl. Phys. {\bf B447} (1995) 35.

\bibitem{harveymoore} J.A. Harvey and G. Moore, Nucl. Phys. {\bf B463}
(1996) 315.

\bibitem{henningson} M. Henningson and G. Moore, {\tt hep-th/9608145}.

\bibitem{kounnas} C. Kounnas and E. Kiritsis, Nucl. Phys. {\bf B442}
(1995) 472; {\sl ibid.} {\bf B456} (1995) 699; 
\\
E. Kiritsis, C. Kounnas, P.M. Petropoulos and J. Rizos, 
{\tt hep-th/9605011}, {\tt hep-th/9608034}. 

\bibitem{yaugroup} P. Candelas, X.C. de la Ossa, P.S. Green and L. Parkes,
Nucl. Phys. {\bf B359} (1991) 21;\\
S. Hosono, A. Klemm, S. Theisen and S.-T. Yau, Comm. Math. Phys.
{\bf 167} (1995) 301.

\bibitem{cardosochoice}
G.L. Cardoso, G. Curio, D. L\"ust and T. Mohaupt, Phys. Lett. {\bf 382B}
(1996) 241.

\bibitem{stsymm}
A. Klemm, W. Lerche and P. Mayr, Phys. Lett. {\bf 357B} (1995) 313.

\bibitem{smallinstanton}
E. Witten, Nucl. Phys. {\bf B460} (1996) 541.

\bibitem{dewitlatest}
B. de Wit, G.L. Cardoso, D. L\"ust, T. Mohaupt and S.-J. Rey,
{\tt hep-th/9607184}.

\bibitem{lustkallosh}
K. Behrndt, G.L. Cardoso, B. de Wit, R. Kallosh, D. L\"ust and
T. Mohaupt, {\tt hep-th/9610105}.



\end{thebibliography}
\end{document}